\documentclass[8.5pt,twoside,twocolumn]{article}

\usepackage[top=2.5cm, bottom=2.5cm, left=2cm , right=2cm]{geometry}
\usepackage[super,sort&compress,comma]{natbib} 
\usepackage{mhchem}
\usepackage{times}
\usepackage{sectsty}
\usepackage{balance} 
\usepackage{graphicx} 
\usepackage{lastpage}
\usepackage[format=plain,singlelinecheck=false,font=small,labelfont=bf,labelsep=space]{caption} 
\usepackage{amssymb}
\usepackage{amsmath}
\usepackage{multirow}
\usepackage{color}
\usepackage{enumitem}
\usepackage{wrapfig}
\usepackage{multicol}
\usepackage{fixltx2e}
\usepackage{dblfloatfix}
\usepackage[squaren,Gray]{SIunits}

\begin{document}
%
%
\twocolumn[
  \begin{@twocolumnfalse}
\noindent\LARGE{\textbf{Surfactant-induced rigidity of interfaces: a unified approach to free and dip-coated films.}}
\vspace{0.6cm}

\noindent\large{\textbf{Lor\`{e}ne Champougny,\textit{$^{a}$} Benoit Scheid,\textit{$^{b\ddag}$} Fr\'{e}d\'{e}ric Restagno,\textit{$^{a}$} Jan Vermant,\textit{$^{c}$} and
Emmanuelle Rio\textit{$^{a}$}$^{\ast}$}}\vspace{0.5cm}

\vspace{0.6cm}

\noindent \normalsize{The behavior of thin liquid films is known to be strongly affected by the presence of surfactants at the interfaces. The detailed mechanism by which the latter enhance film stability is still a matter of debate, in particular concerning the influence of surface elastic effects on the hydrodynamic boundary condition at the liquid/air interfaces. In the present work, ``twin'' hydrodynamic models neglecting surfactant transport to the interfaces are proposed to describe the coating of films onto a solid plate (Landau-Levich-Derjaguin configuration) as well as soap film pulling (Frankel configuration). Experimental data on the entrained film thickness in both configurations can be fitted very well using a single value of the surface elasticity, which is in good agreement with independent measurements by mean of surface expansion experiments in a Langmuir through. The analysis shows how and when the soap films or dip coating experiments may be used to precisely and sensitively measure the surface elasticity of surfactant solutions.}
\vspace{0.5cm}
 \end{@twocolumnfalse}
  ]
%
%
%
%
\footnotetext{\textit{$^{a}$~Laboratoire de Physique des Solides, CNRS \& Universit\'{e} Paris-Sud, 91405 Orsay cedex, France}}
\footnotetext{\textit{$^{b}$~TIPs - Fluid Physics unit, Universit\'{e} Libre de Bruxelles C.P. 165/67, 1050 Brussels, Belgium}}
\footnotetext{\textit{$^{c}$~Department of Materials - ETH Z\"{u}rich, Vladimir-Prelog-Weg 5, 8093 Z\"{u}rich, Switzerland}}
%
%
\footnotetext{\ddag~E-mail: bscheid@ulb.ac.be}
\footnotetext{$^{\ast}$~E-mail: emmanuelle.rio@u-psud.fr}
%
%
%
%
\section{Introduction}
\begin{figure}
\centering
\includegraphics[width=8cm]{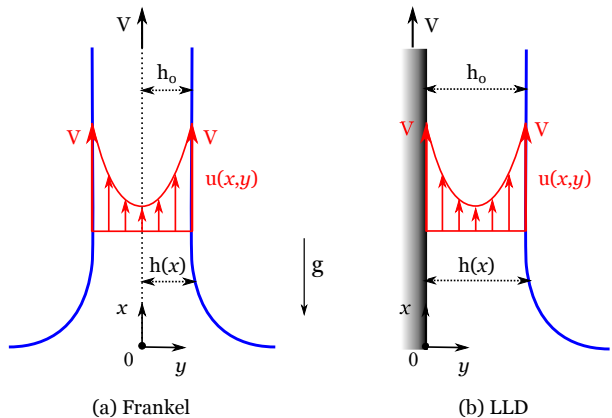}
\caption{Sketches introducing the main notations used in this paper for (a) Frankel configuration, \textit{i.e.} soap film pulling, and (b) LLD configuration, \textit{i.e.} plate coating. The velocity field $u(x,y)$ is defined in the reference frame of the laboratory and is represented in the case of rigid liquid/air interfaces in both Frankel and LLD configurations. Note that $h(x)$ is defined as half the film thickness for a soap film, whereas it stands for the actual thickness of a coated film.}
\label{fig:notations}
\end{figure}

\begin{figure}%
\centering
\includegraphics[width=\linewidth]{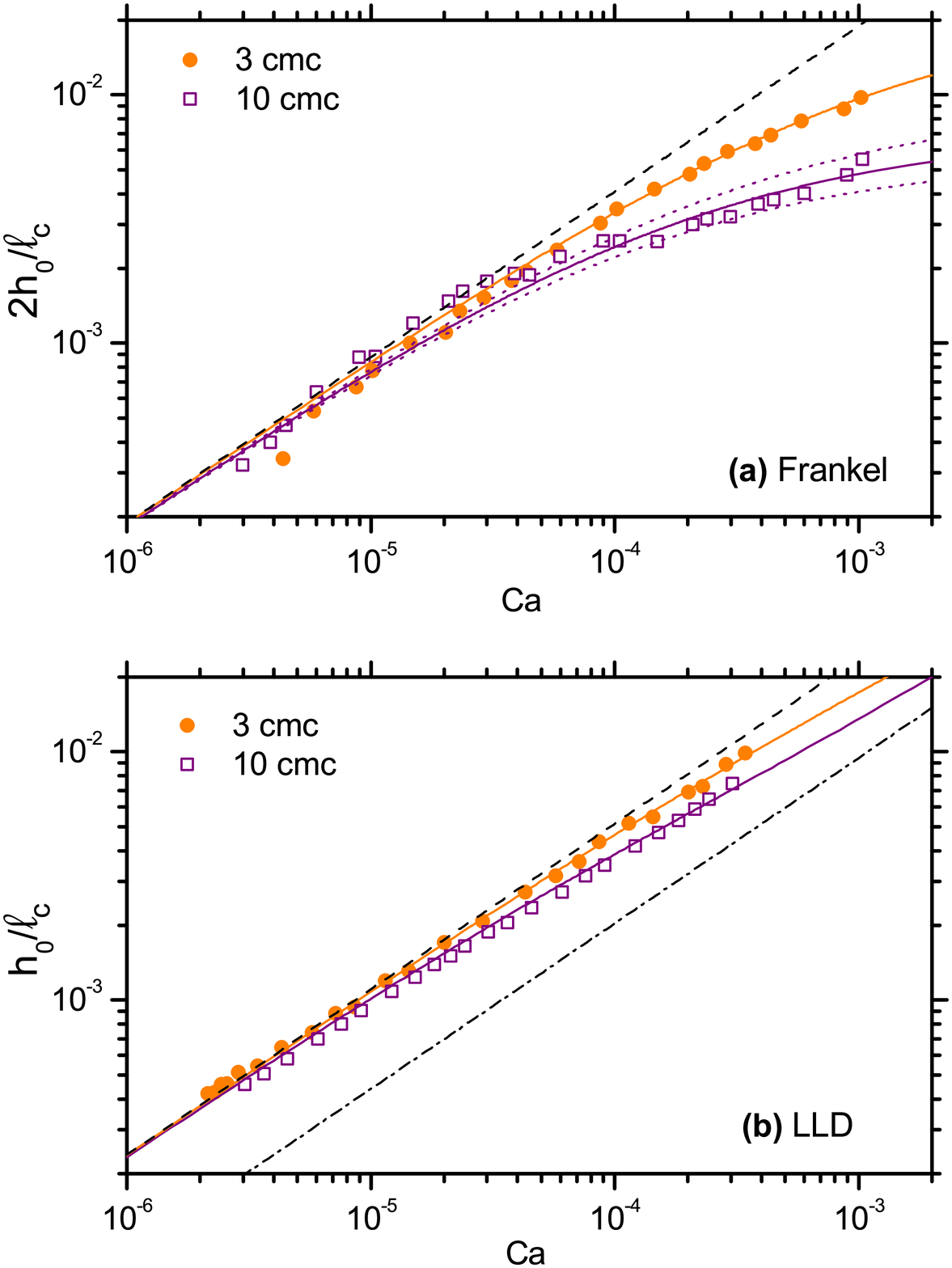}%
\caption{(a) Frankel configuration: the film thickness $2h_0$ normalized by the capillary length $\ell_c$ is plotted as a function of the capillary number $\mathrm{Ca}$. Symbols correspond to experimental data from ref.\cite{Saulnier_phd} for soap films pulled from $\mathrm{C}_{12} \mathrm{E}_6$ solutions of two different concentrations, respectively 3 and 10 times the critical micellar concentration (cmc). The solid lines are fits using the model described in section \ref{sec:model} and the dotted lines give an estimation of the maximal error on the fit. Finally, the dashed line corresponds to Frankel's law (eq. \eqref{eq:scaling} with $K_{\mathrm{Fr}}$), namely the limit of rigid liquid/air interfaces.  --- (b) LLD configuration: same as (a) for coated films entrained from $\mathrm{C}_{12} \mathrm{E}_6$ solutions\cite{Delacotte2012} at the same concentrations as in Frankel configuration. The dashed line corresponds to the LLD law in the limit of a rigid liquid/air interface (eq. \eqref{eq:scaling} with $K_{\mathrm{LLD}}^{\mathrm{rigid}}$) and the dash-dotted line to the limit of a stress-free liquid/air interface (eq. \eqref{eq:scaling} with $K_{\mathrm{LLD}}^{\mathrm{sf}}$).}%
\label{fig:fit_data}%
\end{figure}

%
%
The formation of thin liquid films, either free standing (soap films) or deposited on a solid substrate (coated films), is of utmost importance for many applications. The thickness of soap films within a foam is indeed one of the key ingredients controlling foam destabilization. Understanding the physical and physicochemical parameters that control film thickness is thus necessary when stable foams are required, as in cosmetics or food products for instance, but also if quick foam collapse is sought for\cite{Stevenson2012}. The coating of thin liquid films onto solid substrates is a widespread industrial process, used for surface functionalization as well as surface protection or lubrication\cite{Bhushan1991}. In many cases, the deposition of a liquid layer of controlled thickness is needed. Understanding the influence of physical chemistry on film thickness is all the more crucial as complex liquids, such as colloidal dispersions or emulsions, are often used as coating agents\cite{Kistler1997}. \\
The major difference between free and coated films is that liquid films can be coated onto a solid substrate by viscous entrainment, whereas viscous forces alone are not sufficient to generate a free standing film\cite{vanNieropCORR2009,Seiwert2014}. The additional support that helps pulling a soap film upwards stems from interfacial shear stress, which usually results from spatial gradients in the concentration of surface active agents adsorbed at the liquid/air interfaces\cite{Mysels1959,Bruinsma1992}, but can also be generated by temperature gradients\cite{Scheid2010_thermocapillary,Scheid2012_thermocapillary}. \\
Despite their different fields of applications, the generation of a free standing film and the coating of a liquid film on a plate are similar from the hydrodynamic point of view but have mostly been studied separately in the literature. In particular, no quantitative parallel has been drawn, to our knowledge, between free films and coated films generated from identical surfactant solutions. The first hydrodynamic descriptions of those processes were respectively given by Mysels, Shinoda and Frankel\cite{Mysels1959} for soap films and by Landau, Levich and Derjaguin\cite{Landau1942, Derjaguin1943} for coated films. Both models assume the fluid properties to be independent of position and, using the Stokes equation combined with suitable boundary conditions, they yield the same scaling law for the thickness $h_0$ of a film generated at a constant velocity $V$: 
\begin{equation}
h_0 = K \ell_c \, \mathrm{Ca}^{2/3}.
\label{eq:scaling}
\end{equation}
In the above equation, the thickness $h_0$ is defined as sketched in figure \ref{fig:notations} and $K$ is a numerical prefactor, while $\ell_c = \sqrt{\gamma_0 / \rho g}$ and  $\mathrm{Ca}=\eta \, V / \gamma_0$ are respectively the capillary length and the capillary number, in which $\gamma_0$ is the equilibrium surface tension, $\rho$ is the liquid density, $g$ is the gravitational acceleration and $\eta$ is the liquid bulk viscosity. This scaling law follows from a balance between viscous shear entrainment and capillary suction and should be valid as long as gravitational drainage can be neglected, \textit{i.e.} for $\mathrm{Ca}^{1/3} \ll 1$. The value of the prefactor $K$ in eq. \eqref{eq:scaling} essentially depends on the film boundary conditions at the liquid/air interface. In the Landau-Levich-Derjaguin (LLD) configuration with a \emph{pure} liquid, a no slip condition is taken at the solid/liquid interface while the liquid/air interface is simply assumed to be stress free. However, when pulling a free standing film out of a solution containing surface active agents, the presence of sustaining interfacial stresses must be accounted for in the boundary condition at the liquid/air interfaces\cite{Sagis2011}. This is the reason why, in their original work, Mysels, Shinoda and Frankel assumed that the interfaces behave as if entrained without slip by a solid wall; such boundary condition is here referred to as ``rigid''. \\
The hydrodynamic hypothesis of rigid liquid/air interfaces was found in good qualitative agreement with many experiments for both soap films\cite{Bruinsma1992, LiontiAddad1992, Lal1994, Adelizzi2004, Berg2005, Saulnier2011} and coated films\cite{Morey1940, Krechetnikov2005, Snoeijer2008} at low capillary numbers ($\mathrm{Ca} < 10^{-5}$). It turned out to hold as well in other geometries, like fluid coating on fibers\cite{Quere1999, Shen2002} or bubbles moving in tubes\cite{Bretherton1961}. Nevertheless, the dispersion of data points from one experiment to another\cite{vanNierop2008}, as well as the deviations from eq. \eqref{eq:scaling} observed at higher capillary numbers ($\mathrm{Ca} \sim 10^{-5}-10^{-4}$), where the condition $\mathrm{Ca}^{1/3} \ll 1$ still holds\cite{LiontiAddad1992, vanNierop2009_formation, Saulnier2011, Delacotte2012}, remain unexplained. \\
%
%
To go beyond the assumption of rigid interfaces, several models have been proposed to account for the elastic and/or viscous behavior of the liquid/air interfaces for both LLD\cite{Park1991, Tiwari2006, Campana2010, Campana2011} and Frankel\cite{Schwartz1999, Naire2001, Seiwert2014} configurations. Because the mechanical response of the interfaces is dictated by the interfacial dynamics of surfactants, most of the models make the assumption of water-insoluble surfactants, for which the complex adsorption-desorption dynamics can be disregarded. Still, the surface rheological parameters involved in the models -- surface elasticity and viscosity -- are difficult to measure experimentally. Quantitative comparison between thin film models including surface rheology and experimental data is scarce in the literature\cite{Scheid2010_LLD, Seiwert2014, Bhamla2014}. Moreover, most experimental data available on thin films were obtained with water-soluble surfactant: comparison to models neglecting mass exchange between the surface and the bulk is thus to be done with caution -- but remains possible\cite{Sonin1994} in certain conditions, as will be discussed later on.\\
Delacotte \textit{et al.}\cite{Delacotte2012} have measured the thickness of the film coated on a solid plate as a function of the plate velocity for the non-ionic surfactant $\mathrm{C}_{12} \mathrm{E}_6$ at various concentrations, and their data are reported in figure \ref{fig:fit_data}a. Similar measurements have been performed by Saulnier \textit{et al.}\cite{Saulnier2011} on soap films, using the same surfactant and concentrations as Delacotte \textit{et al.}, and their data are plotted in figure \ref{fig:fit_data}b. At high capillary numbers, both studies observe a deviation from the $\mathrm{Ca}^{2/3}$ behavior, as pedicted by eq. \eqref{eq:scaling}, towards thinner films. They suggest a qualitative explanation based on the idea, developed by Prins \textit{et al.}\cite{Prins1967}, Lucassen \textit{et al.}\cite{Lucassen1981} and later by Qu\'{e}r\'{e}\cite{Quere1999}, that the ability of the liquid/air interfaces of a film to sustain surface tension gradients decreases with film thickness. \\
Building on the works of Park\cite{Park1991} and Seiwert \textit{et al.}\cite{Seiwert2014}, we propose to rationalize quantitatively the deviations from Frankel and LLD laws observed at high capillary numbers in a unified description including the surface elasticity $E$. This parameter, which quantifies the compressibility of an interface, is defined as\cite{Gibbs1906}
%
%
\begin{equation}
E = A \, \frac{\partial \gamma}{\partial A},
\label{eq:elasticite}
\end{equation}
where $A$ is the area of the liquid/air interface and $\gamma$ is the surface tension. \\
%
In section \ref{sec:scalings}, the different regimes exhibited by the data in figure \ref{fig:fit_data} and the corresponding transitions are discussed in terms of scaling laws. ``Twin'' models, coupling the hydrodynamics and the dynamics of insoluble surfactants at the liquid/air interfaces, are then recalled in section \ref{sec:model} and numerically solved for soap films as well as for coated films. Quantitative comparison between models and experiments on the nonionic sufactant $\mathrm{C}_{12} \mathrm{E}_6$ is performed in section \ref{resolution_and_fit}: we find that the data can be well described assuming the surface elasticity is independent on the pulling velocity. In section \ref{Langmuir_trough}, we finally report on Langmuir trough measurements of the surface elasticity of $\mathrm{C}_{12} \mathrm{E}_6$ solutions, which turn out to be in good agreement with the values extracted from Frankel and LLD experiments.
%
%
\section{Scaling analysis \label{sec:scalings}}
%
%
\subsection{The limiting cases of rigid and stress-free interfaces}
%
%
The typical lengthscales in the $x$- and $y$-directions (defined in figure \ref{fig:notations}) are respectively given by the lengthscale $\ell$ of the transition region -- called the dynamic meniscus -- which connects the static meniscus to the flat part of the film, and the thickness $h_0$ of the film far from the liquid bath. The lengthscale $\ell \sim \sqrt{h_0 \ell_c}$ is deduced from the scaling analysis of the curvature matching condition between the dynamic and static meniscii:
\begin{equation}
\partial_{xx} h = \frac{\sqrt{2}}{\ell_c} \qquad \text{for} \quad x \rightarrow - \infty
\label{eq:matching}
\end{equation}
and is typically of the order of $20-200~\mathrm{\mu m}$ in the range of $\mathrm{Ca}$ explored experimentally ($\mathrm{Ca}=10^{-6}-10^{-3}$). In this paper, we use the notation $\partial_i f$ to denote the derivative of the function $f$ with respect to the spatial coordinate $i$. If the pulling velocity $V$ is used to scale the velocity and the pressure $P$ is assumed to be uniform across the film, like the capillary pressure in the static meniscus $\gamma_0/\ell_c$, eq. \eqref{eq:scaling} can be recovered (with $K=1$) from the scaling analysis of the $x$-component of Stokes equations in the absence of gravity:
\begin{equation}
\eta \, \partial_{yy} u =  \partial_x P.
\label{eq:NSx}
\end{equation}
The value of the prefactor $K$ in eq. \eqref{eq:scaling} is obtained from the asymptotic matching eq. \eqref{eq:matching}. In the LLD geometry, the stress-free boundary condition at the liquid/air interface yields\cite{Landau1942}
\begin{equation}
K_{\mathrm{LLD}}^{\mathrm{sf}} = 0.9458,
\label{eq:K_LLD_fluid}
\end{equation}
which is in good agreement with experimental data for pure liquids\cite{Snoeijer2008,Delacotte2012}. Because it assumes rigid liquid/air interfaces, the Mysels-Shinoda-Frankel model -- which we henceforth call ``Frankel's model'' since it is referred to as such in the literature -- boils down to the LLD model with a factor $2$, the liquid/air interfaces being viewed as two solid walls and the stress along the vertical symmetry axis of the free film being zero. In our notations (see figure \ref{fig:notations}), the prefactor in eq. \eqref{eq:scaling} is still $K_{\mathrm{LLD}}^{\mathrm{sf}} = K_{\mathrm{Fr}}$ but the actual thickness of the soap film is twice the thickness of a pure liquid LLD film. \\
Finally, in the LLD geometry, the limiting case of a rigid liquid/air interface changes the prefactor in eq. \eqref{eq:scaling} into 
\begin{equation}
K_{\mathrm{LLD}}^{\mathrm{rigid}} = 4^{2/3} K_{\mathrm{LLD}}^{\mathrm{sf}}.
\label{eq:K_LLD_rigid}
\end{equation}
Note that $K_{\mathrm{LLD}}^{\mathrm{rigid}}$ is larger than $2 K_{\mathrm{Fr}}$, since the capillary suction that opposes the entrainment is caused by the presence of only one meniscus in the LLD case but two meniscii in the Frankel case.
%
%
\subsection{Partially rigid interfaces}
%
As soon as we want to refine the boundary condition at the liquid/air interface, \textit{i.e.} to explore intermediate cases between the stress-free and rigid limits, the balance of tangential forces at the liquid/air interface has to be considered:
\begin{equation}
\left. \eta \, \partial_{y} u \right|_{y=h(x)} = \partial_x \gamma.
\label{eq:tangent}
\end{equation}
This equation shows that spatial variations of the surface tension generate a tangential stress, called Marangoni stress, at the liquid/air interfaces. In the following, we will assume that the surface tension gradient $\partial_x \gamma$ scales like $E/\ell$. This is equivalent to saying that Marangoni stresses are locally controlled by the surface elasticity $E$ defined by eq. \eqref{eq:elasticite}.\\
The viscous stress at the interface $\left. \eta \, \partial_{y} u \right|_{y=h(x)}$ can be computed by integrating \eqref{eq:NSx} along $y$ between $0$ and $h(x)$, so that eq. \eqref{eq:tangent} becomes
\begin{equation}
\left. \eta \, \partial_{y} u \right|_{y=0} + h \, \partial_x P = \partial_x \gamma,
\label{eq:NS_int}
\end{equation}
where the film thickness $h$ (or half the film thickness in the case of a soap film) scales like $h_0$.
%
%
%
%
\subsection{Frankel configuration}
%
%
%
\begin{figure}[t]
\centering
\includegraphics[width=8cm]{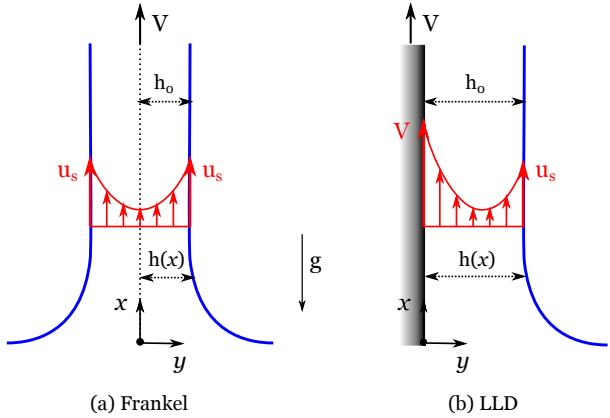}
\caption{Sketches of the vertical velocity field $u(x,y)$ for (a) Frankel configuration and (b) LLD configuration in the case of \emph{partially} rigid liquid/air interfaces. The interfacial velocity $u_{\mathrm{s}}(x)$ is introduced to account for finite surface elasticity. All velocities are defined in the reference frame of the laboratory.}
\label{fig:notations_b}
\end{figure}
When considering a film with partially rigid interfaces, it is no longer obvious that the vertical velocity $u$ scales like $V$, since the interfacial velocity $u_{\mathrm{s}}$ introduces a second velocity scale, which is \textit{a priori} different from $V$, as pictured in figure \ref{fig:notations_b}a. However, thanks to the symmetry of the film with respect to the vertical axis\cite{Howell1996}, $\left. \partial_{y} u \right|_{y=0}$ vanishes in eq. \eqref{eq:NS_int} in the case of Frankel configuration, leading to the velocity-independent equation
%
\begin{equation}
\partial_x \gamma = h \, \partial_x P,
\end{equation}
where the left-hand side represents the Marangoni stress dragging liquid up, while the right-hand side is the capillary suction driving liquid down. From this equation we can draw a velocity-independent scaling law for $h_0$:
\begin{equation}
h_0 \sim \ell_c \, \frac{E}{\gamma_0}.
\label{eq:scaling_Fr_b}
\end{equation}
In the case of partially rigid interfaces, the thickness of a soap film is thus found to scale independently of the pulling velocity. Such a velocity-independent regime was already identified by Scheid \textit{et al.}\cite{Scheid2010_thermocapillary} in the case of surface tension gradients induced by thermocapillary effects. It was also obtained in the context of thin films pulled from surfactant-polymer mixtures by Bruinsma \textit{et al.}\cite{Bruinsma1992}. The transition between this velocity-independent regime, corresponding to partially rigid interfaces, and Frankel's regime, corresponding to rigid interfaces, is found by comparing eq. \eqref{eq:scaling_Fr_b} to the scaling of Frankel's law (eq. \eqref{eq:scaling}) and occurs for
\begin{equation}
\mathrm{Ca}^{\ast} \sim \left( \frac{E}{\gamma_0} \right)^{3/2}.
\label{eq:transition}
\end{equation}
Experimental data (figure \ref{fig:fit_data}a) show that low capillary numbers ($\mathrm{Ca} < \mathrm{Ca}^{\ast}$) correspond to Frankel's regime. On the contrary, the film thickness seems to converge, without reaching it, towards a velocity-independent regime at high capillary numbers ($\mathrm{Ca} \gg \mathrm{Ca}^{\ast}$). Figure \ref{fig:schemas_scalings}a summarizes this scaling analysis for Frankel configuration. An order of magnitude for the critical capillary number at which the transition occurs can be estimated from figure \ref{fig:fit_data}a: for $\mathrm{C}_{12} \mathrm{E}_6$ solutions at concentrations $3$ and $10$ cmc, we find $\mathrm{Ca}^{\ast} \sim 10^{-4}$. \\
This brings evidence that the ``rigidity'' of liquid/air interfaces cannot be considered as a property of the surfactant solution. We indeed show that, at a given position on the film, the mechanical behaviour of the liquid/air interfaces of a free film can range from rigid (\textit{i.e.} well described by Frankel's law) to partially rigid upon increasing the pulling velocity. A similiar conclusion had already been reached by Stebe \textit{et al.}\cite{Stebe1995} in a Bretherton-like configuration, namely for gas bubbles moving in a capillary tube.
%
%
%
\subsection{LLD configuration}
\begin{figure}[t]
\centering
\includegraphics[width=7.5cm]{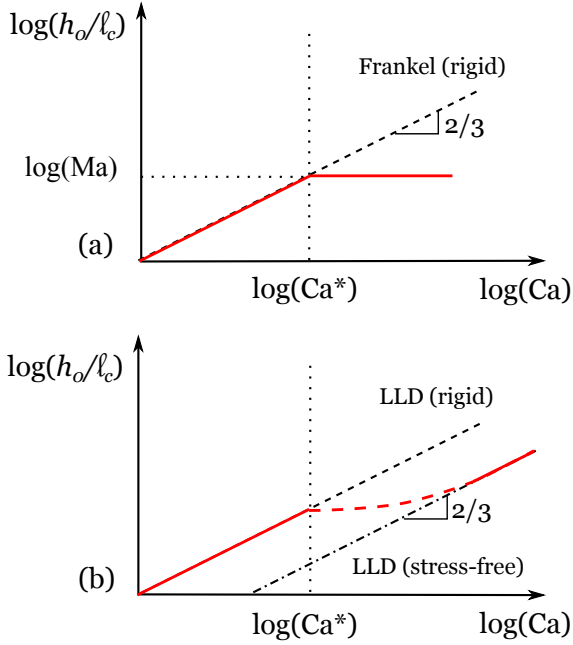}%
\caption{Log-log diagrams of the capillary number dependence of film thickness for (a) Frankel configuration and (b) LLD configuration, obtained by scaling analysis. (a) For soap films, a velocity-independent regime is expected at high capillary numbers ($\mathrm{Ca} > \mathrm{Ca}^{\ast}$ given by eq. \eqref{eq:transition}), whereas Frankel's power law \eqref{eq:scaling} remains valid at lower capillary numbers ($\mathrm{Ca} < \mathrm{Ca}^{\ast}$). (b) For coated films, the limit of the LLD model with \emph{stress-free} interfaces is obtained at high capillary numbers $\mathrm{Ca} \gg \mathrm{Ca}^{\ast}$. A transition regime, which cannot be simply described in terms of a scaling law, is expected for $\mathrm{Ca} \gtrsim \mathrm{Ca}^{\ast}$. Finally, the LLD model with \emph{rigid} interfaces should be recovered for $\mathrm{Ca} < \mathrm{Ca}^{\ast}$.}%
\label{fig:schemas_scalings}%
\end{figure}

%
%
In the LLD configuration, the symmetry of the film with respect to the vertical axis is lost and the $\left. \partial_{y} u \right|_{y=0}$ term in eq. \eqref{eq:NS_int} must be kept. However, the no-slip condition at the solid/liquid interface ensures that, in the case of a partially rigid (or even stress-free) boundary condition at the liquid/air interface, $V$ remains the right velocity scale for $u$, as pictured in figure \ref{fig:notations_b}b. In the end, using the scaling of eq. \eqref{eq:matching} to compute $\ell$, all terms in eq. \eqref{eq:NS_int} can be estimated:
\begin{equation}
\underbrace{-\eta \, \left. \partial_{y} u \right|_{y=0}}_{\sim \, \frac{\eta V}{h_0}}%
+ \underbrace{\partial_x \gamma}_{\sim \, \frac{E}{\sqrt{\ell_c h_0}}}%
= \underbrace{h(x) \, \partial_x P}_{\sim \, \gamma_0 \left( \frac{h_0}{\ell_c^3} \right)^{1/2}}.
\label{eq:scaling_LLD_b}
\end{equation}
In the limit of a partially rigid liquid/air interface ($u_{\mathrm{s}}<V$), equation \eqref{eq:scaling_LLD_b} expresses the balance between the forces that pull the film upwards (left-hand terms) and the capillary suction, which drives liquid downwards (right-hand term). The left-hand terms of eq. \eqref{eq:scaling_LLD_b} account for two contributions: the viscous stress associated to the film entrainment and the Marangoni stress associated to surface tension gradients. For ``large'' capillary numbers, the viscous stress dominates and we recover the classical LLD law in the \emph{stress-free} limit, namely with prefactor $K_{\mathrm{LLD}}^{\mathrm{sf}}$ and $u_{\mathrm{s}} \lesssim V$. Upon decreasing the capillary number, the Marangoni stress can become of the order of the viscous one. This intermediate regime, which occurs around the critical capillary number $\mathrm{Ca}^{\ast}$ given by eq. \eqref{eq:transition}, cannot be simply described by a scaling law. For even lower capillary numbers, we expect to observe a transition towards the \emph{rigid} limit, dominated by the Marangoni stress, namely LLD law with prefactor $K_{\mathrm{LLD}}^{\mathrm{rigid}} = 4^{2/3} K_{\mathrm{LLD}}^{\mathrm{sf}}$ and $u_{\mathrm{s}}=V$. \\
The stress-free and rigid limits, as well as the transition, are sketched in figure \ref{fig:schemas_scalings}b, although both limits have never been observed together experimentally for a given surfactant concentration. For a given surfactant concentration, it is indeed the value of surface elasticity that determines which portion of the curve sketched in figure \ref{fig:schemas_scalings}b is observed in the range of capillary numbers where our hypotheses (no gravitational drainage, no inertia) are valid.
%
%
%
\section{Hydrodynamic model including surface elasticity\label{sec:model}}
%
%
\subsection{Assumption of water-insoluble surfactants}
%
%
The typical timescale $\tau$ in Frankel and LLD experiments can be defined as the time spent by surfactants in the dynamic meniscus, where the surface is stretched, namely $\tau = \ell / V$. This timescale, which lies in the range $3-600~\mathrm{ms}$ for the experimental data discussed here, is to be compared to the characteristic adsorption time of surfactants at the liquid/air interface, denoted $\tau_{\mathrm{ads}}$. \\
For the non-ionic surfactant $\mathrm{C}_{12} \mathrm{E}_6$, the adsorption is purely diffusion-limited\cite{Bendure1971} and $\tau_{\mathrm{ads}}$ is found experimentally\cite{Lin2002} to be of the order of $100~\mathrm{s}$ for a bulk concentration $c=10~\mathrm{cmc}$. We thus have $\tau \ll \tau_{\mathrm{ads}}$ for $\mathrm{C}_{12} \mathrm{E}_6$ for the whole range of capillary numbers probed, hence a negligible contribution of adsorption to the surface concentration of surfactants $\Gamma$. This supports the assumption of insoluble surfactants to describe the water-soluble surfactant $\mathrm{C}_{12} \mathrm{E}_6$, at least in the present context of ``rapid'' film formation. In this case, the general definition of the surface elasticity (eq. \eqref{eq:elasticite}) becomes
\begin{equation}
E_{\mathrm{insol}} = - \Gamma \, \frac{\partial \gamma}{\partial \Gamma}.
\label{eq:E_insoluble}
\end{equation}
In the following, we make the further assumption that $E_{\mathrm{insol}}$ is constant, so that eq. \eqref{eq:E_insoluble} can be integrated to find the equation of state
\begin{equation}
\gamma (\Gamma) = \gamma_0 - E_{\mathrm{insol}} \ln \left( \frac{\Gamma}{\Gamma_0} \right),
\label{eq:EOS}
\end{equation}
with $\gamma_0 = \gamma (\Gamma_0)$ the surface tension for the interface in equilibrium with the bulk, \textit{i.e.} far from the film, and $\Gamma_0$ the corresponding surface concentration. Note that this equation of state differs from the one used by Park\cite{Park1991} and Seiwert \textit{et al.}\cite{Seiwert2014}, which was the linearized equation of state obtained when assuming that $\Gamma$ stays close to $\Gamma_0$. The results presented in figure \ref{fig:Delta_Gamma_gamma} will show that this hypothesis is valid only at small capillary numbers. Detailed comparison between the solutions obtained using the linearized and non-linearized equations of state is discussed in appendix \ref{NL_vs_lin}.
%
%
\subsection{Hypotheses and governing equations}
%
%
%
Similar to the derivation of Frankel's and LLD laws, we assume stationarity and the existence of a region far from the liquid bath where the film thickness is uniform and equals to $h_0$. Surface diffusion is neglected and all equations are written at leading order in the frame of the lubrication approximation, \textit{i.e.} $h_0/\ell \ll 1$, corresponding to small interfacial slopes ($\partial_{x}h \ll 1$). Under those assumptions, the film thickness $h(x)$ for coated films, or half the film thickness for soap films, the vertical velocity $u(x,y)$, the vertical surface velocity $u_{\mathrm{s}}(x) = u(x,h(x))$, the pressure $P(x,y)$ and the surface concentration of surfactants $\Gamma(x)$ obey the following system of coupled differential equations: 
\begin{itemize}
	\item Lubrication equations:
	\begin{equation}
	\begin{aligned}
	\partial_x P &= \eta \, \partial_{yy} u, \\
	\partial_y P &= 0,
	\end{aligned}
	\label{eq:NS_dim}
	\end{equation}
	\item Normal force balance at the liquid/air interface:
	\begin{equation}
	\left. P \right|_{y=h(x)}= P_0 - \gamma \, \partial_{xx} h,
	\label{eq:Laplace_dim}
	\end{equation}
	\item Tangential force balance at the liquid/air interface:
	\begin{equation}
	\eta \, \left. \partial_{y} u \right|_{y=h(x)} = \partial_x \gamma,
	\label{eq:tangent_dim}
	\end{equation}
	\item Mass conservation at the liquid/air interface:
	\begin{equation}
	\partial_x (u_{\mathrm{s}} \Gamma) = 0,
	\label{eq:surf_cons_dim}
	\end{equation}
	\item Mass conservation in the bulk of the film:
	\begin{equation}
	\partial_x (\bar{u} h) = 0,
	\label{eq:mass_cons_dim}
	\end{equation}
	where 
	\begin{equation}
	\bar{u}(x) = \frac{1}{h(x)} \int_0^{h(x)} u(x,y) \mathrm{d}y,
	\label{eq:vitesse_moy}
	\end{equation}
	is the average velocity in the film (or half the film for a soap film) at a given height $x$. 
\end{itemize}
We refer the interested reader to references\cite{Park1991, Seiwert2014} for the details of the derivation in the LLD and Frankel configurations, respectively. The main difference with these references in the present model is the use of the nonlinear equation of state \eqref{eq:EOS} in eq. \eqref{eq:Laplace_dim}, as well as in eq. \eqref{eq:tangent_dim} to account for the correction to the capillary pressure gradient due to the variation of surface tension, which was in fact accounted for in ref.\cite{Park1991} but not in ref.\cite{Seiwert2014}. The influence of these choices on the results will be discussed later in the text and in appendix \ref{NL_vs_lin}.
%
%
%
%
\subsection{Non-dimensionalization and boundary conditions}
%
%
The pressure and velocity fields are eliminated from the previous system using either the symmetry condition $\left. \partial_y u \right|_{y=0}=0$ for the Frankel configuration or the no-slip condition at the solid wall $\left. u \right|_{y=0}=V$ for the LLD configuration, which leaves three one-dimensional fields $h(x)$, $u_{\mathrm{s}}(x)$ and $\Gamma(x)$ to be determined. \\
The equations are non-dimensionalized as follows: $y$ and $h$ are rescaled by $h_0$, $x$ by $\ell$, $\gamma$ by $\gamma_0$, $u_{\mathrm{s}}$ by $V$ and $\Gamma$ by $\Gamma_0$. We also define the aspect ratio $\epsilon = h_0/\ell \ll 1$, which scales the interfacial slope $\partial_x h$. Considering that the dominant balance is capillary suction versus viscous entrainment, \textit{i.e.} the scaling of $h_0$ is still given by Frankel's law \eqref{eq:scaling}, leads to $\epsilon = \mathrm{Ca}^{1/3}$, as discussed in reference\cite{Landau1942}. \\
To solve the problem, we take $x=0$ a position in the flat region of the film (far from the liquid bath) and $x=-L$ the position where the asymptotic matching between the dynamic and static meniscii is made. Note that the solutions do not depend on $L$ as long as it is large enough, so that the curvature $h^{\prime \prime}(-L)$ has converged towards a constant value, denoted $h^{\prime \prime}(-\infty)$. The non-dimensionalized boundary conditions then read: $h(0) = 1$, $h^{\prime} (0) = 0$, $h^{\prime \prime} (0) = a$, $u_{\mathrm{s}}(0) = 1$ and $\Gamma (-L) = 1$, with $a$ a small parameter (here $a = 10^{-3}$). Finally, the non-dimensionalized system of equations to be solved are given below for each configuration, using a prime to denote $x$-derivatives.

%
\subsection{Frankel configuration \label{Frankel_system}}
%
%
%
\begin{figure}[t]
\centering
\includegraphics[width=\linewidth]{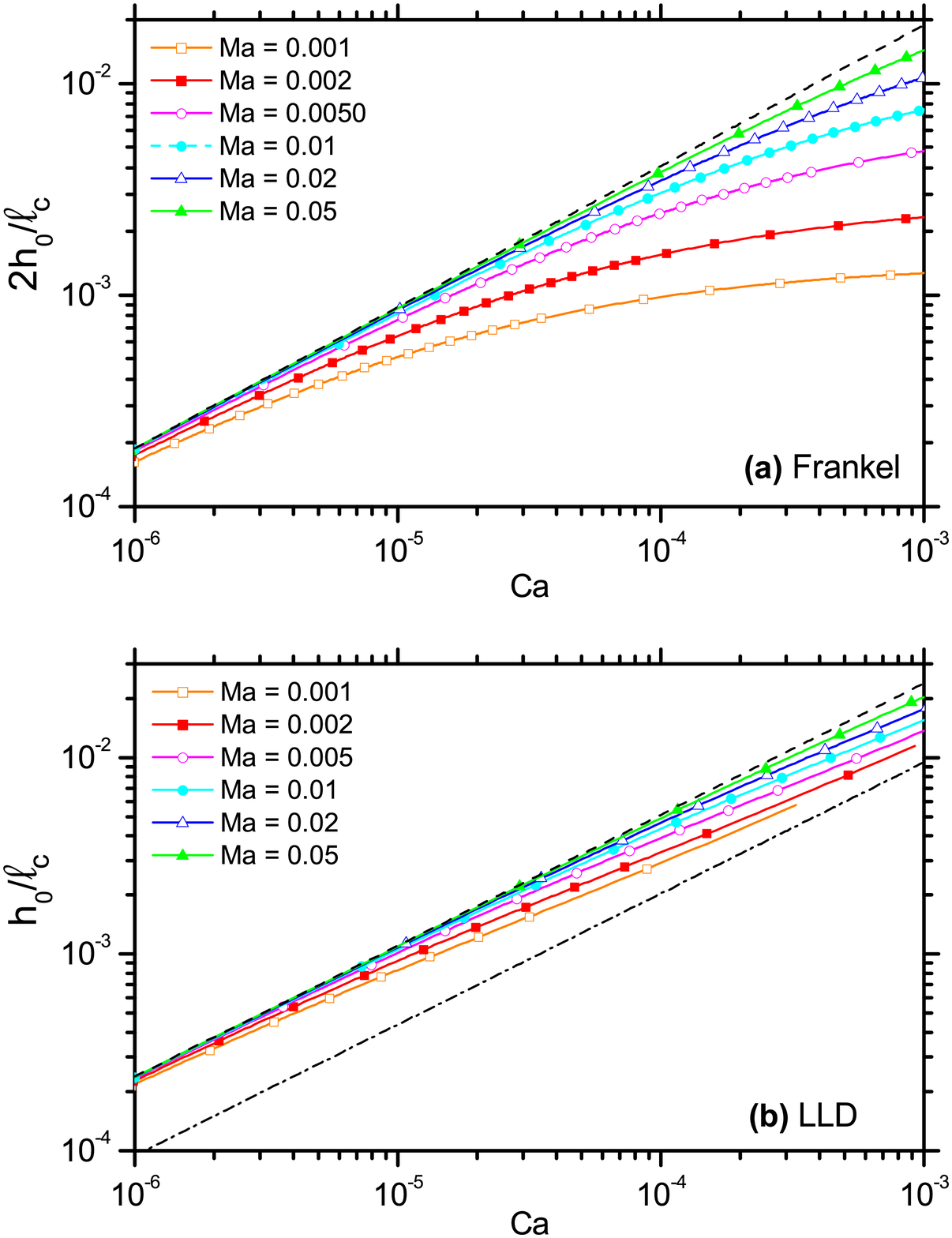}%
\caption{(a) Frankel configuration: the predicted film thickness $2h_0$ normalized by the capillary length $\ell_c$ is plotted as a function of the capillary number $\mathrm{Ca}$. Different symbols/colors correspond to different values of the Marangoni number $\mathrm{Ma}$. The dashed line corresponds to Frankel's law, namely the limit of rigid liquid/air interfaces.  --- (b) LLD configuration: same as (a) for the coated film thickness $h_0$. The dashed line corresponds to the LLD law for the limit of a rigid liquid/air interface and the dash-dotted line to the limit of a stress-free liquid/air interface.}%
\label{fig:comp_courbes_theo}%
\end{figure}
\begin{itemize}
	\item Differential equation for the film thickness $h(x)$:
	\begin{equation}
	 h^{\prime \prime \prime} = 3 \frac{(1 - u_{\mathrm{s}} h)}{\gamma h^3} + \mathrm{Ma} \frac{h^{\prime \prime}}{\gamma} \frac{\Gamma^{\prime}}{\Gamma} \approx 3 \frac{(1 - u_{\mathrm{s}} h)}{\gamma h^3},
	\label{eq:diff_h_Fr}
	\end{equation}
	\item Differential equation for the surface velocity $u_{\mathrm{s}}(x)$:
	\begin{equation}
	u_{\mathrm{s}}^{\prime} = - \frac{3}{\Lambda} \frac{u_{\mathrm{s}}}{h^2}(1 - u_{\mathrm{s}} h),
	\label{eq:diff_us_Fr}
	\end{equation}
	\item Differential equation for the surface concentration $\Gamma(x)$:
	\begin{equation}
	\Gamma^{\prime} = -\frac{\Gamma}{u_{\mathrm{s}}} u_{\mathrm{s}}^{\prime},
	\label{eq:diffGamma}
	\end{equation}
	\item Equation of state
	\begin{equation}
	\gamma (\Gamma) = 1 - \mathrm{Ma} \ln (\Gamma) \approx 1,
	\label{eq:EOS_ndim}
	\end{equation}
\end{itemize}
where we have defined the Marangoni number as
\begin{equation}
\mathrm{Ma} = \frac{E_{\mathrm{insol}}}{\gamma_0}
\label{eq:Ma}
\end{equation}
and the dimensionless parameter $\Lambda$ as 
\begin{equation}
\Lambda = \frac{\mathrm{Ma}}{\mathrm{Ca}^{2/3}}.
\label{eq:Lambda}
\end{equation}
The above system of equations is solved under the assumption that $\mathrm{Ma} \ll 1$, \textit{i.e.} neglecting the variation of surface tension around its equilibrium value in equations \eqref{eq:diff_h_Fr} and \eqref{eq:EOS_ndim} for the whole range of $\mathrm{Ca}$ probed experimentally. We shall justify this assumption in appendix \ref{NL_vs_lin}, where we compare the solution obtained by solving the whole system to the one derived from the approximated system. In the case of soap films, we show that the above approximation remains correct as long as $\mathrm{Ma} \lesssim 0.1$. \\
In the end, this leaves us with a single parameter, $\Lambda$, that controls the solutions of the system and that we call the \emph{rigidity} parameter. Indeed the critical capillary number $\mathrm{Ca}^{\ast}$ identified in the scaling analysis (section \ref{sec:scalings}) corresponds to $\Lambda = 1$. This brings evidence that $\Lambda$ is the natural parameter to describe the transition between the limit of rigid liquid/air interfaces, recovered for $\Lambda \gg 1$, and the regime of partially rigid liquid/air interfaces, obtained for $\Lambda \lesssim 1$.
%
%
\subsection{LLD configuration \label{LLD_system}}
%
%
\begin{itemize}
	\item Differential equation for the film thickness $h(x)$:
	\begin{equation}
	 h^{\prime \prime \prime} = 6 \frac{[2 - (1+u_{\mathrm{s}}) h]}{\gamma h^3} + \mathrm{Ma} \frac{h^{\prime \prime}}{\gamma} \frac{\Gamma^{\prime}}{\Gamma} \approx 6 \frac{[2 - (1+u_{\mathrm{s}}) h]}{\gamma h^3},
	\label{eq:diff_h_LLD}
	\end{equation}
	\item Differential equation for the surface velocity $u_{\mathrm{s}}(x)$:
	\begin{equation}
	u_{\mathrm{s}}^{\prime} = - \frac{2}{\Lambda} \frac{u_{\mathrm{s}}}{h^2}[3 - (1+2u_{\mathrm{s}}) h],
	\label{eq:diff_us_LLD}
	\end{equation}
	\item Differential equation for the surface concentration: eq. \eqref{eq:diffGamma},
	\item Equation of state: eq. \eqref{eq:EOS_ndim}.
\end{itemize}
The above system is solved under the assumption that $\mathrm{Ma} \ll 1$, so that the Marangoni term in equations \eqref{eq:diff_h_LLD} and \eqref{eq:EOS_ndim} can be neglected. In the case of coated films, we show in appendix \ref{NL_vs_lin} that this approximation remains correct as long as $\mathrm{Ma} \lesssim 0.01$. As in Frankel configuration, we are left with a single control parameter $\Lambda$, defined in eq. \eqref{eq:Lambda}.
%
%
\section{Resolution of the model and fit of experimental data\label{resolution_and_fit}}
%
%
\subsection{Computing film thickness versus capillary number\label{computing_h}}
%
%
%
The systems of equations described in subsections \ref{Frankel_system} and \ref{LLD_system} are solved numerically using the continuation software AUTO-07p\cite{AUTO07P}, for different values of the dimensionless parameter $\Lambda$. For a given $\Lambda$ and a given $\mathrm{Ma}$, we compute the film thickness $h_0$ (or half the film thickness in the case of a soap film) in the flat part of the film using the curvature matching condition eq. \eqref{eq:matching} in its dimensionless form (for $x \rightarrow - \infty$):
\begin{equation}
\frac{h_0}{\ell_c} = \frac{h^{\prime \prime}(-\infty)}{\sqrt{2}} \mathrm{Ca}^{2/3}.
\label{eq:matching_ndim}
\end{equation}
Setting a numerical value for the Marangoni number $\mathrm{Ma}$, we are able to deduce $h_0$ as a function of the capillary number $\mathrm{Ca}=(\mathrm{Ma}/\Lambda)^{3/2}$ from the mastercurve $h_0(\Lambda)$. For Frankel (resp. LLD) configuration, we thus have a family of theoretical curves $h_0(\mathrm{Ca})$ parametrized by $\mathrm{Ma}$, as illustrated in figure \ref{fig:comp_courbes_theo}a (resp. \ref{fig:comp_courbes_theo}b). For both configurations, the theoretical curves $h_0(\mathrm{Ca})$ superimpose on the rigid limit (dashed line) at ``low'' capillary numbers and then deviate from it around the critical capillary number $\mathrm{Ca}^{\ast}$ that increases with $\mathrm{Ma}$. All these features are consistent with the qualitative behavior predicted from the scaling analysis (figure \ref{fig:schemas_scalings}). Note that, in the case of soap films (figure \ref{fig:comp_courbes_theo}a), no velocity-independent regime is truly reached at ``high'' capillary numbers in the range of $\mathrm{Ca}$ and $\mathrm{Ma}$ relevant for experiments. Additionally, in the case of coated films (figure \ref{fig:comp_courbes_theo}b), the stress-free limit itself is not recovered at ``high'' capillary numbers in the relevant range of parameters (see also appendix \ref{NL_vs_lin}).
%
%
%
\subsection{From experimental data to surface elasticity \label{extract_E}}
%
%
%
\vspace{-5mm}
\begin{multicols}{2}
\begin{figure*}[b]
\centering
\includegraphics[width=13cm]{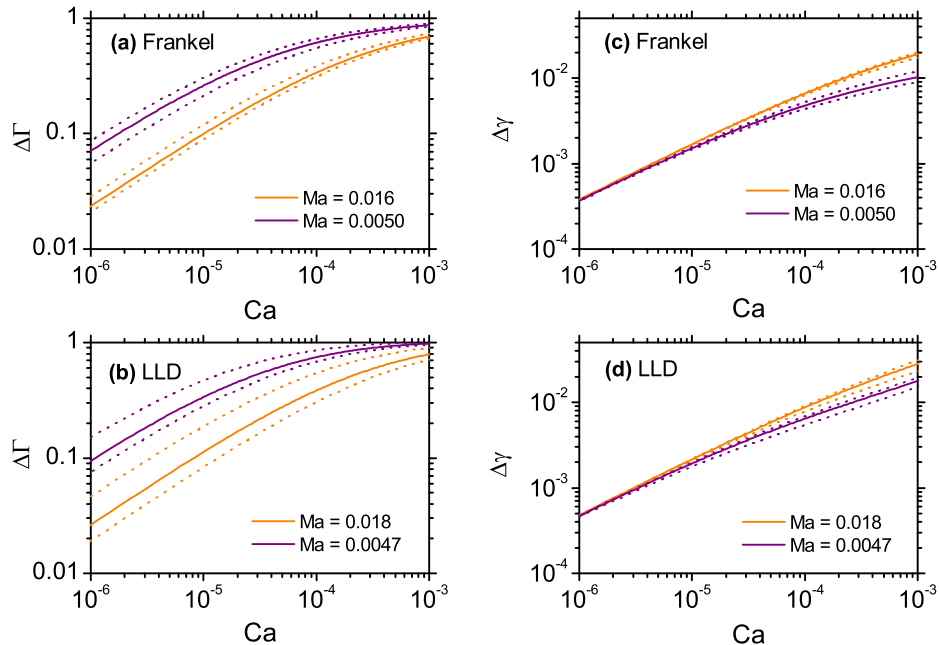}%
\caption{The dimensionless surface concentration difference $\Delta \Gamma$ and the dimensionless surface tension difference $\Delta \gamma$ between the flat part of the film and the liquid bath are plotted as functions of the capillary number $\mathrm{Ca}$ for (a), (c) Frankel configuration and (b), (d) LLD configuration. In all cases, the Marangoni number $\mathrm{Ma}$ has been set to the value extracted from the fit of the corresponding experimental data (see table \ref{tab_C12E6}). The dotted lines stand for the confidence intervals on $\Delta \gamma$ and $\Delta \Gamma$ deduced from the confidence intervals on the value of surface elasticity (see table \ref{tab_C12E6}).}
\label{fig:Delta_Gamma_gamma}%
\end{figure*}
\end{multicols}
\begin{table}
	\centering
		\begin{tabular}{cccc}
			\multicolumn{2}{c}{$\mathrm{C}_{12} \mathrm{E}_6$} & 3 cmc & 10 cmc  \\
			\hline 
			\multirow{2}{*}{Frankel}	& $\mathrm{Ma} \times 10^3$ & $16 \, \substack{+2 \\ -3}$ & $5.0 \, \substack{+1.5 \\ -1.0}$ \\
			 & $E_{\mathrm{insol}}~(\mathrm{mN/m})$ & $0.54 \, \substack{+0.08 \\ -0.09}$ & $0.17 \, \substack{+0.05 \\ -0.03}$ \\
			\hline
			\multirow{2}{*}{LLD}	& $\mathrm{Ma} \times 10^3$ & $18 \, \substack{+7 \\ -8}$ & $4.7 \, \substack{+1.3 \\ -1.9}$ \\		
			
			 & $E_{\mathrm{insol}}~(\mathrm{mN/m})$ & $0.59 \, \substack{+0.21 \\ -0.27}$ & $0.15 \, \substack{+0.04 \\ -0.06}$ \\
		\end{tabular}
		\caption{Summary of the fitted Marangoni numbers $\mathrm{Ma}$ obtained for free standing and coated films pulled from $\mathrm{C}_{12} \mathrm{E}_6$ solutions of concentration 3 cmc and 10 cmc (see figure \ref{fig:fit_data}). The corresponding surface elasticity in the insoluble limit $E_{\mathrm{insol}} = \gamma_0 \, \mathrm{Ma}$ was then deduced taking the experimental values for surface tension $\gamma_0 = 34.5~\mathrm{mN/m}$ at $22~\celsius$ for Frankel configuration\cite{Saulnier2011} and $\gamma_0 = 32.3~\mathrm{mN/m}$ at $25~\celsius$ for LLD configuration\cite{Delacotte2012}.}
		\label{tab_C12E6}
\end{table}
For practical use, a non-linear fit of the mastercurve $h_0(\Lambda)$ is proposed in appendix \ref{NL_fit} in both Frankel and LLD configurations. We can then obtain an analytical expression of $h_0$ as a function of the capillary number $\mathrm{Ca}=(\mathrm{Ma}/\Lambda)^{3/2}$, where $\mathrm{Ma}$ is the only adjustable parameter. This expression is used to fit the experimental data available on free-standing (resp. coated) films pulled from $\mathrm{C}_{12} \mathrm{E}_6$ solutions (symbols in figure \ref{fig:fit_data}) and extract a value for $\mathrm{Ma}$. The best fits are shown in figure \ref{fig:fit_data} (solid lines) for two different surfactant concentrations above the critical micellar concentration (cmc). The corresponding values for the Marangoni number $\mathrm{Ma}$ and surface elasticity in the insoluble limit $E_{\mathrm{insol}}$ are summarized in table \ref{tab_C12E6}. The confidence interval on the Marangoni number is estimated from the upper and lower values associated to ``reasonable'' fits of the data, as illustrated by the dotted lines in figure \ref{fig:fit_data}a for $10$ cmc.\\
The main result of table \ref{tab_C12E6} is that the experimental data for both Frankel and LLD configurations can be rationalized by a single value of surface elasticity for a given surfactant concentration. This brings evidence that surface elasticity in the insoluble limit is an inherent property of a surfactant solution (\textit{e.g.} independent of the pulling velocity and on the configuration) and is thus also relevant for the description of thin films stabilized by water-soluble surfactants.
\subsection{Surface tension gradient \label{results2}}
%
%
For the particular values of $\mathrm{Ma}$ displayed in table \ref{tab_C12E6}, the (dimensionless) relative surface concentration variation between the liquid bath and the flat part of the film, defined as $\Delta \Gamma =1-\Gamma(x=0)$, is computed as a function of the capillary number $\mathrm{Ca}$, for Frankel (figure \ref{fig:Delta_Gamma_gamma}a) and LLD (figure \ref{fig:Delta_Gamma_gamma}b) configurations. The relative surface concentration difference $\Delta \Gamma$ between the flat part of the film and the liquid bath increases with the capillary number, and eventually becomes of the order of unity. This shows that the linearization of the equation of state \eqref{eq:EOS} is no longer justified for $\mathrm{Ca} \gtrsim 10^{-4}$ and introduces biases that are discussed in appendix \ref{NL_vs_lin}. \\
Similarly, the (dimensionless) relative surface tension variation between the liquid bath and the flat part of the film, defined as $\Delta \gamma =\gamma(x=0)-1$ is deduced from the equation of state eq. \eqref{eq:EOS_ndim} and plotted as a function of the capillary number $\mathrm{Ca}$ for Frankel (figure \ref{fig:Delta_Gamma_gamma}c) and LLD (figure \ref{fig:Delta_Gamma_gamma}d) configurations, for the particular values of $\mathrm{Ma}$ given in table \ref{tab_C12E6}. The relative surface tension difference $\Delta \gamma$ between the flat part of the film and the liquid bath increases with the capillary number, but does not reach more than a few percents in the range of capillary numbers probed experimentally, supporting the neglect of the Marangoni terms in eqs. \eqref{eq:diff_h_Fr}, \eqref{eq:EOS_ndim} and \eqref{eq:diff_h_LLD}. \\
Numerically, this corresponds to a surface tension difference $\Delta \gamma \approx 0.2~\mathrm{mN/m}$ for a soap film pulled from a $\mathrm{C}_{12} \mathrm{E}_6$ solution at 3 cmc at a typical velocity $V \approx 2~\mathrm{mm/s}$ ($\mathrm{Ma} \approx 0.018$, $\mathrm{Ca} \approx 6 \times 10^{-5}$). This order of magnitude is consistent with theoretical results obtained by Seiwert \textit{et al.}\cite{Seiwert2014} for Frankel configuration. Combining the present study to the recent predictions by Saulnier \textit{et al.}\cite{Saulnier2014} for surface tension gradients in the upper non-stationary part of soap films, it is now possible to forecast the surface tension difference across the whole film. For a $\mathrm{C}_{12} \mathrm{E}_6$ solution at 3 cmc and $\mathrm{Ca} \approx 6 \times 10^{-5}$, this total surface tension difference is found of the order of $1~\mathrm{mN/m}$, which is small compared to the surface tension of usual surfactant solutions, but well in line with recent experimental work on surface tension gradients in soap films by Caps \textit{et al.}\cite{Caps2013}.
%
%
%
\section{Langmuir trough measurements of surface elasticity \label{Langmuir_trough}}
%
%
To our knowledge, direct measurements of surface elasticity in thin films, such as Prins's ones\cite{Prins1967} for a solution of SDS at a concentration of 3 times the cmc, do not exist for $\mathrm{C}_{12} \mathrm{E}_6$. The surface elasticities of $\mathrm{C}_{12} \mathrm{E}_6$ solutions have been measured using a dynamic drop tensiometer\cite{Lucassen2001}, but only up to bulk concentrations of 1 cmc. In this section, we present surface elasticity measurements in a Langmuir trough for $\mathrm{C}_{12} \mathrm{E}_6$ at concentrations above the cmc, to be compared to the values computed independently in subsection \ref{extract_E}.
%
%
\subsection{Materials and methods}
%
%
\begin{figure}[t]
\centering
\includegraphics[width=\linewidth]{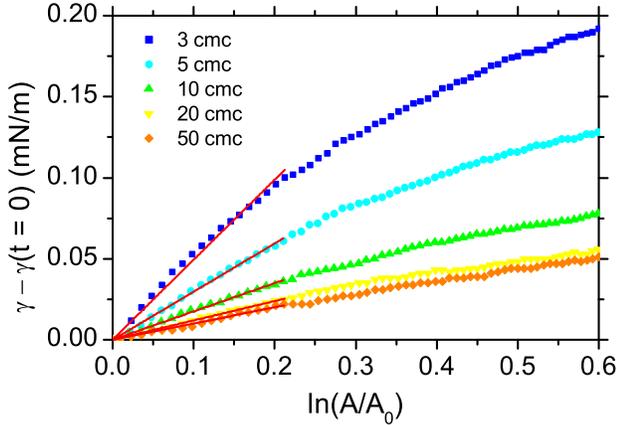}%
\caption{The dynamic surface tension is plotted as a function of the area between the barriers during their \emph{expansion} at a velocity $v=30~\mathrm{mm/min}$, for $\mathrm{C}_{12} \mathrm{E}_6$ solutions at various concentrations. The red lines correspond to linear fits of the data using eq. \ref{eq:elasticite_integree} in the area range corresponding to a barrier displacement of $1~\mathrm{cm}$. The slope directly gives the surface elasticity $E$.}%
\label{fig:fits_Langmuir}%
\end{figure}
%
%
Five solutions, of concentrations 3, 5, 10, 20 and 50 times the cmc ($\mathrm{cmc} = 0.07~\mathrm{mM}$), are prepared diluting hexaethylene glycol monododecyl ether ($\mathrm{C}_{12} \mathrm{E}_6$, purchased from Sigma Aldrich) in ultrapure (twice-distilled) water. When changing solution, the Langmuir trough and its barriers are thoroughly rinced with ethanol and ultrapure water, and the platinium Wilhelmy plate used for surface tension measurments is rinced with ethanol and ultrapure water and finally put to the flame. The whole setup is enclosed in a box, within which the atmosphere is humidified up to about $80~\%$ in order to reduce evaporation from the Langmuir trough. The trough itself is kept at a constant temperature of $23~\celsius$ by a circulating thermostatic bath. \\
An experiment is carried out as follows: the barriers are initially at a given position, enclosing an area $A_0=A(t=0)$. At time $t=0$, the barriers are set into motion at a constant velocity $v$, expanding the area $A(t)$ between them, up to the user-set maximum position, where they stop. The surface tension $\gamma(t)$ is recorded during the whole experiment, using the Wilhelmy plate technique. In our experiments, the velocity $v$ can be varied from $1$ to $120~\mathrm{mm/min}$, corresponding to capillary numbers $\mathrm{Ca}= \eta v /\gamma(t=0)$ in the range $5\times 10^{-7} - 6\times 10^{-5}$.
%
%
\subsection{Results and discussion}
%
%
%
\begin{figure}[t]
\centering
\includegraphics[width=\linewidth]{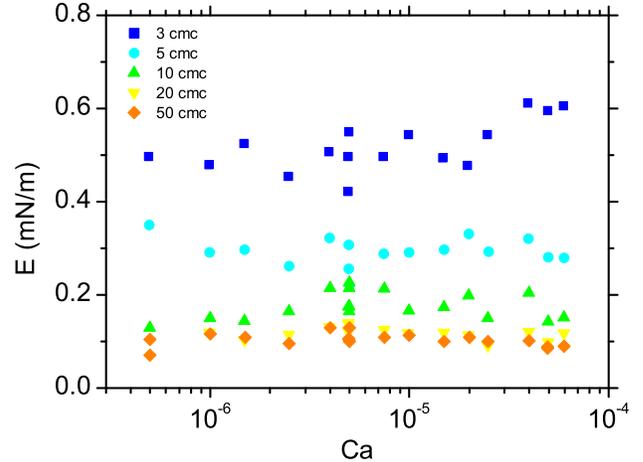}%
\caption{The surface elasticity measured in the Langmuir trough is plotted versus the capillary number $\mathrm{Ca}= \eta v /\gamma(t=0)$ for different $\mathrm{C}_{12} \mathrm{E}_6$ solutions.}%
\label{fig:elasticites_Langmuir}%
\end{figure}
\begin{figure}[t]
\centering
\includegraphics[width=\linewidth]{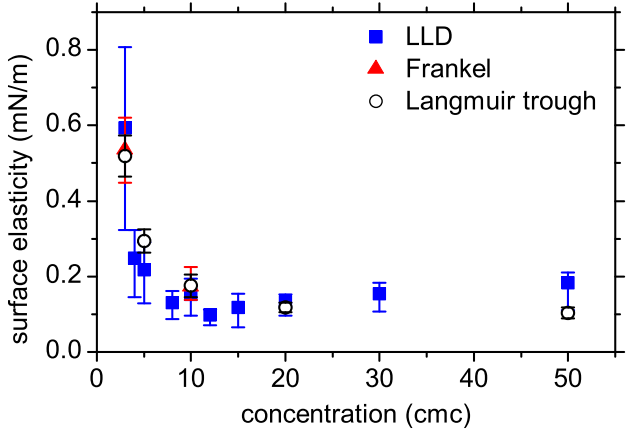}%
\caption{The surface elasticity is plotted as a function of surfactant concentration (counted in number of cmc) for the non-ionic surfactant $\mathrm{C}_{12} \mathrm{E}_6$. The squares and triangles are the values of $E_{\mathrm{insol}}$ obtained by fitting the corresponding experimental curves $h_0 (\mathrm{Ca})$, either in Frankel\cite{Saulnier_phd} or in LLD\cite{Delacotte2012} configuration, using the model described in section \ref{LLD_system}. The error bars stand for the maximal error on $E_{\mathrm{insol}}$ deduced from the confidence interval on the Marangoni number (details can be found in the caption of table \ref{tab_C12E6}). The empty circles correspond to independent measurements of the surface elasticity $E$ performed in a Langmuir trough, averaged over different barrier velocities. The error bars then stand for the standard deviation.}%
\label{fig:elasticites}%
\end{figure}
Figure \ref{fig:fits_Langmuir} shows the measured surface tension minus its initial value $\gamma(t=0)$ as a function of the logarithm of the area between the barriers, normalised by its initial value $A_0$, for $v=30~\mathrm{mm/min}$ and different $\mathrm{C}_{12} \mathrm{E}_6$ bulk concentrations. Under the assumption of a constant surface elasticity $E$, one would expect the curves shown in figure \ref{fig:fits_Langmuir} to follow
\begin{equation}
\gamma - \gamma(t=0) = E \ln \left( \frac{A}{A_0}\right),
\label{eq:elasticite_integree}
\end{equation}
that is to say, to be straight lines. \\
Let us estimate the amount of surface dilation involved in film pulling experiments. The area $A-A_0$ created in the dynamic meniscus, where surface stretching occurs, is proportional to the surface velocity difference $\Delta u_s$ across the dynamic meniscus multiplied by the time $\ell/V$ spent by fluid elements in the dynamic meniscus. Considering that the initial area $A_0$ is the area of the static meniscus and taking into account that $\ell \sim \ell_c \mathrm{Ca}^{1/3}$, we end up with
\begin{equation}
\left( \frac{A}{A_0} \right)_{\mathrm{exp}} \sim 1+\frac{\Delta u_s}{V} \mathrm{Ca}^{1/3}.
\label{eq:stretching_films}
\end{equation}
The surface velocity difference $\Delta u_s$ can be computed from figure \ref{fig:usinf} and eq. \eqref{eq:stretching_films} leads to $\ln[(A/A_0)_{\mathrm{exp}}] < 0.1$ for $\mathrm{Ca}$ in the range $10^{-6}-10^{-3}$and $\mathrm{Ma}\sim 0.01$. This supports the linear fit of the data by equation \eqref{eq:elasticite_integree} (solid red lines in figure \ref{fig:fits_Langmuir}) for small values of the area change, despite the global non-linear variation of $\gamma-\gamma(t=0)$ with $\ln (A/A_0)$. \\
Repeating the experiment for different barrier velocities $v$, we are able to compute the surface elasticity $E$ as a function of the capillary number $\mathrm{Ca}$, as shown in figure \ref{fig:elasticites_Langmuir}. The surface elasticity turns out to be independent of the capillary number in the velocity range probed in the experiments, allowing to consider the average elasticity $E$ for each surfactant concentration. The averaged surface elasticity can then be compared to the values fitted from Frankel\cite{Saulnier2011,Saulnier_phd} and LLD\cite{Delacotte2012} experiments, as described in section \ref{resolution_and_fit}, for the same surfactant concentrations. The comparison is shown in figure \ref{fig:elasticites}. \\
The values of the surface elasticity $E$ measured in a Langmuir trough, where the surfactant reservoir can be considered infinite, are in very good agreement with the values of $E_{\mathrm{insol}}$ extracted from experiments on thin films. This result may be surprising since Prins \textit{et al.}\cite{Prins1967} showed that the surface elasticity decreases when the film thickness increases, which has been interpreted as a reservoir effect\cite{Quere1999}. The local surfactant concentration in thin films is thought to be much lower than the bulk concentration: since they lack surfactant molecules to populate a newly-created interface, thin films should sustain surface tension gradients better than thicker films do. Our data however indicate that this reservoir effect do not contribute significantly to the value of the surface elasticity in thin films stabilized by $\mathrm{C}_{12} \mathrm{E}_6$. \\
Instead, we attribute the dependence of the elasticity with the bulk concentration  observed both in the LLD and Frankel experiments (table \ref{tab_C12E6}) and the Langmuir experiments (figure \ref{fig:elasticites_Langmuir}) to the transport of surfactants brought by advection from the bulk to the newly created surface. This advective transport is a pure consequence of the incompressibility of the liquid. Therefore, a stretched surface will contain more surfactants if the solution underneath is more concentrated, which in turn has the effect to decrease the elasticity, as it is observed in figure \ref{fig:elasticites}, even though only in the range between 3 and 10 cmc. Above 10 cmc, the elasticity remains constant, and thus does not depend on advection transport anymore. The sensitivity of the elasticity with respect to bulk concentration indicates that the surface excess is not equal to the surface concentration, as it is usually assumed for diluted systems\cite{Chang1995}. Finally, the fact that this concentration dependence is independent on the configuration -- i.e. trough, LLD or Frankel -- means that the elasticity is an inherent property of the surfactant solutions, at least in the range of parameters that have been explored in the present work and for deformations that are faster than the inverse of the adsoprtion time.
\section{Conclusion}
We have presented two consistent hydrodynamical models including surface elasticity, for soap film pulling and plate coating experiments, respectively. The predictions of the models, assuming that mass exchange with the bulk can be neglected, are compared to experimental data available for soap and coated films stabilized by the water-soluble surfactant $\mathrm{C}_{12} \mathrm{E}_6$ in order to extract a value of the surface elasticity. For a given surfactant concentration, both Frankel and LLD configurations yield the same value of the surface elasticity, which is confirmed quantitatively by independent measurements in a Langmuir trough. The latter show that the surface elastic behavior of the interface is dominated by the surface elasticity which drives local Marangoni gradients, whereas the contribution of reservoir effects is negligible. Hence coating or film pulling experiments can be used in practice, together with the results of the models presented in this article (see appendix \ref{NL_fit}), to measure the surface elasticity of surfactant solutions, at least in the case of slowly adsorbing surfactants. Additionally, we were able to predict the surface tension difference between the liquid bath and the flat part of the film as a function of the capillary number.\\
As a consequence of our analysis, it is found that the ``rigidity'' of a liquid/air interface, \textit{i.e.} the nature of the boundary condition at this interface, \emph{cannot} be considered as a property of the surfactant solution. Saulnier \textit{et al.}\cite{Saulnier2014} already demonstrated that the upper part of the film is well described by a stress-free boundary condition, whereas its lower part obeys Frankel's law for $\mathrm{Ca} < \mathrm{Ca}^{\ast}$, showing that rigid interfaces are a good approximation in that zone. Similarly, we have shown here that, for a given solution, the pulling velocity dependency of the film thickness is rationalized only when considering that the mechanical behaviour of the liquid/air interfaces ranges from totally to partially rigid upon increasing the pulling velocity. This change in the boundary condition stems from a change in the relative contributions of viscous and elastic entrainments to the force balance, and is controlled by the parameter $\Lambda=\mathrm{Ma}/\mathrm{Ca}^{2/3}$. Our work thus shows that the paradigm of a $2/3$ power law in $\mathrm{Ca}$, used to fit experimental film thickness for both Frankel and LLD configurations by many different authors and for decades, might well be obsolete in the case of films containing surface active agents. \\
Although we have been focusing here on a plane geometry, the work presented in this paper could in principle be extended to other geometries, like fiber coating\cite{OuRamdane1997, Quere1999, Shen2002} or bubbles moving in capillaries\cite{Bretherton1961}, provided the curvature of the static meniscus $\sqrt{2}/ \ell_c$ appearing in the matching condition \eqref{eq:matching} is replaced by $1/r$, where $r$ is either the fiber radius or the bubble cap radius. 
%
%
\appendix
%
\section{Comparison of linearized and non-linearized equations of state \label{NL_vs_lin}}
%
%
\begin{figure}[t]
\centering
\includegraphics[width=\linewidth]{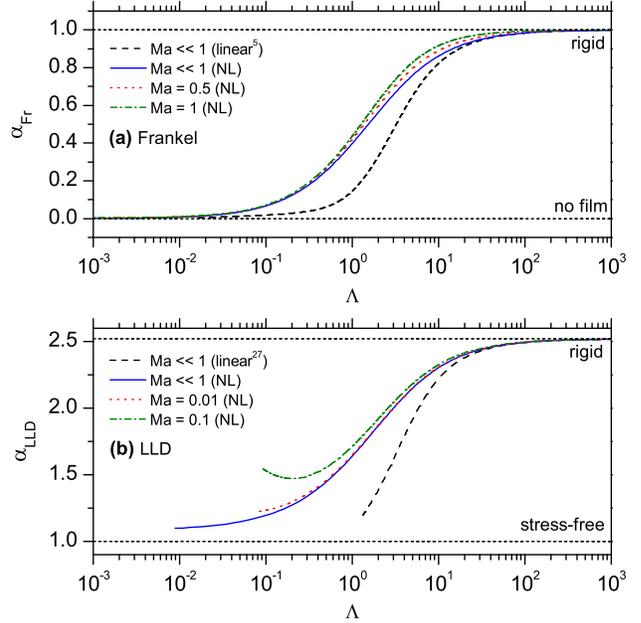}%
\caption{The thickening factor $\alpha$ is plotted as a function of the dimensionless parameter $\Lambda$ for (a) Frankel configuration and (b) LLD configuration. The different curves were obtained under different assumptions: \emph{non-linear} equation of state and $\mathrm{Ma} \ll 1$ (solid lines), \emph{linear} equation of state and $\mathrm{Ma} \ll 1$ (long-dashed lines), non-linear equation of state and finite values for $\mathrm{Ma}$ (dotted and dot-dashed lines). Note that, for a typical Marangoni number $\mathrm{Ma} \sim 10^{-2}$ (as obtained from the experimental data) and capillary numbers between $10^{-6}$ and $10^{-3}$, the control parameter $\Lambda$ ranges from $1$ to $10^2$.}%
\label{fig:annexe}%
\end{figure}
\begin{figure}[t]
\centering
\includegraphics[width=\linewidth]{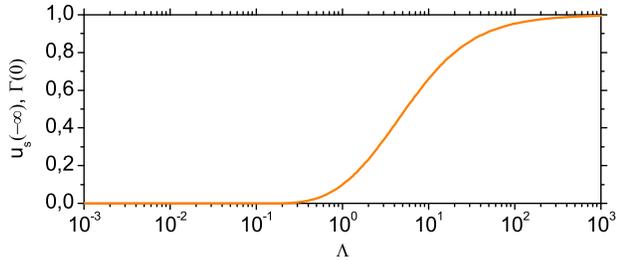}%
\caption{The surface velocity in the static meniscus $u_{\mathrm{s}}(-\infty)$ is plotted as a function of the control parameter $\Lambda$. Note that $u_{\mathrm{s}}(-\infty)$ is equal to the surface concentration in the flat part of the film $\Gamma(0)$ thanks to the surfactant number conservation eq. \eqref{eq:surf_cons_dim}.}%
\label{fig:usinf}%
\end{figure}

In this section, we discuss some assumptions we have made in the models presented in sections \ref{Frankel_system} and \ref{LLD_system}. Unlike in references\cite{Park1991, Seiwert2014}, we have kept the non-linearized form of the equation of state \eqref{eq:EOS} instead of using its linearized version:
\begin{equation}
\gamma (\Gamma) = \gamma_0 - E_{\mathrm{insol}} \, \frac{\Gamma - \Gamma_0}{\Gamma_0}.
\label{eq:EOS_lin}
\end{equation}
The main consequence of this choice is that the surface tension gradient writes $\partial_x \gamma = -E_{\mathrm{insol}} \, \Gamma^{\prime}/\Gamma$ in our case, instead of $\partial_x \gamma = -E_{\mathrm{insol}} \, \Gamma^{\prime}/\Gamma_0$ in the linear case. Using the conservation of surfactants at the interface eq. \eqref{eq:diff_us_Fr}, the surface tension gradient can be expressed independently of the surface concentration as $\partial_x \gamma = E_{\mathrm{insol}} \, u_{\mathrm{s}}^{\prime}/u_{\mathrm{s}}$ in the non-linear case and $\partial_x \gamma = E_{\mathrm{insol}} \, u_{\mathrm{s}}^{\ast} \, u_{\mathrm{s}}^{\prime}/u_{\mathrm{s}}^2$ in the linear case, where the surface velocity $u_{\mathrm{s}}^{\ast}=u_s(-\infty)$ at the junction between the dynamic and static meniscii needs to be determined \textit{a posteriori}. \\
Still under the assumption $\mathrm{Ma} \ll 1$ in equations \eqref{eq:diff_h_Fr} and \eqref{eq:diff_h_LLD}, which shall be precised in the last paragraph of this section, we compute the thickening factor $\alpha_{\mathrm{Fr}}$ (resp. $\alpha_{\mathrm{LLD}}$) in Frankel configuration (resp. in LLD configuration) defined as
\begin{equation}
\alpha_{\mathrm{Fr}} = \frac{h^{\prime \prime}(-\infty)}{\sqrt{2} \, K_{\mathrm{Fr}}} \qquad \left( \text{resp.} \quad \alpha_{\mathrm{LLD}} = \frac{h^{\prime \prime}(-\infty)}{\sqrt{2} \, K_{\mathrm{LLD}}^{\mathrm{sf}}} \right),
\label{eq:def_alpha}
\end{equation}
as a function of the control parameter $\Lambda$. Comparison to equation \eqref{eq:matching_ndim} shows that the thickening factor $\alpha_{\mathrm{Fr}}$ (resp. $\alpha_{\mathrm{LLD}}$) is simply the ratio between the film thickness predicted by our model and the one predicted by the Frankel (resp. LLD) model. \\
The function $\alpha_{\mathrm{Fr}}(\Lambda)$ (resp. $\alpha_{\mathrm{LLD}}(\Lambda)$) is computed using the \emph{linearized} equation of state eq. \eqref{eq:EOS_lin} (dashed line) and compared to our solution with the non-linear equation of state (solid line) in figure \ref{fig:annexe}a (resp. \ref{fig:annexe}b). Note the dashed line in figure \ref{fig:annexe}a (resp. \ref{fig:annexe}b) had previouly been obtained by ref.\cite{Seiwert2014} (resp. ref.\cite{Park1991}) for soap film pulling and plate coating respectively. For both configurations, we observe the following discrepencies between the results obtained with the linear and non-linear equations of state: 
\begin{enumerate}[label=\roman*)]
	\item The linearized equation of state gives a sharper transition between the rigid and stress-free behaviors: the transitions only spans over two decades, instead of three for the non-linearized equation of state,
	\item Using the linearized equation of state can lead to significant underestimation of the film thickness (up to a factor of 3 for $\Lambda \sim 1$ in the case of soap films), and thus on the deduced surface elasticity,
	\item In Frankel configuration, the curve $h_0(\mathrm{Ca})$ reconstructed using the linearized equation of state (\textit{i.e.} from the dashed line in figure \ref{fig:annexe}a) goes through a maximum. On the contrary, the reconstruction using the non-linear equation of state (\textit{i.e.} from the solid line in figure \ref{fig:annexe}a) yields a monotonous increase of the thickness with $\mathrm{Ca}$.
%
%
\end{enumerate}
For Frankel configuration, the fully rigid case (Frankel's model) is recovered at large $\Lambda$, namely at large Marangoni number or low capillary number, no matter the equation of state. For low $\Lambda$, the thickness tends to zero: in practice, the interfacial stresses are no longer sufficient to pull a stable film, hence the ``no film'' limit in figure \ref{fig:annexe}a (lower short-dashed line). \\
For LLD configuration, the maximal thickening factor $4^{2/3} \approx 2.52$ corresponding to the rigid limit is also obtained at large $\Lambda$ for both the linear and non-linear equation of state. However, the stress-free limit $\alpha_{\mathrm{LLD}}=1$ of the LLD model is not recovered for small values of $\Lambda$, neither for the linearized nor for the non-linearized equation of state. This is likely due to a loss of convergence in our numerical calculations when decreasing $\Lambda$, and is associated with the impossibility for the surface velocity to become negative in the case of \emph{insoluble} surfactants. As illustrated in figure \ref{fig:usinf} for LLD configuration, the concentration in the flat film $\Gamma(0)$ tends to zero when decreasing $\Lambda$, and so does the surface velocity in the static meniscus $u_{\mathrm{s}}(-\infty)$, since they are related by the conservation law eq. \eqref{eq:surf_cons_dim}. On the contrary, as shown by Stebe \textit{et al.}\cite{Stebe1995}, considering \emph{soluble} surfactants modifies eq. \eqref{eq:surf_cons_dim} and makes it possible to reach negative surface velocities, and thus to recover the stress-free limit. The curves displayed in figure \ref{fig:annexe} are therefore stopped when the surface velocity in the static meniscus becomes zero and this limitation is intrinsic to the insoluble description of surfactants. Note that the position of the stopping point is very sensitive to the model used and that our simplified model (solid curves) provides the widest range of converged solutions. \medskip \\
The Marangoni term in equations \eqref{eq:diff_h_Fr}, \eqref{eq:EOS_ndim} (subsection \ref{Frankel_system}) and \eqref{eq:diff_h_LLD} (subsection \ref{LLD_system}) has been neglected under the assumption $\mathrm{Ma} \ll 1$. Figure \ref{fig:annexe} compares the curves $\alpha_{\mathrm{Fr}}(\Lambda)$ and $\alpha_{\mathrm{LLD}}(\Lambda)$ obtained by solving the system including the Marangoni terms (dotted and dot-dashed lines) to the ones obtained by solving the approximated system (solid lines). All predictions including the Marangoni terms fall on top of the approximated solution for $\mathrm{Ma} \lesssim 0.1$ in the case of soap films and for $\mathrm{Ma} \lesssim 0.01$ in the case of coated films. Consequently, the assumption $\mathrm{Ma} \ll 1$ does not have the same meaning in Frankel and LLD configurations, but remains valid in both cases given the values of $\mathrm{Ma}$ extracted from the fit of the experimental data (see table \ref{tab_C12E6}).
%
%
\section{Non-linear fit of $h_0$ vs $\mathrm{Ca}$ \label{NL_fit}}
%
\subsection{Frankel configuration}
For practical use, we here propose a non-linear fit of the thickening factor $\alpha_{\mathrm{Fr}}$ defined as $h_0$ divided by the half film thickness predicted by Frankel's model, namely $K_{\mathrm{Fr}} \, \ell_c \, \mathrm{Ca}^{2/3}$:
\begin{equation}
\alpha_{\mathrm{Fr}}(\Lambda) \approx \frac{1}{2} \left[1 - \tanh \left(186.3 - 186.1 \, \Lambda^{0.0026} \right) \right].
\label{eq:_alpha_Lambda_Frankel}
\end{equation}
The theoretical half film thickness as a function of the capillary number is then recovered for a given Marangoni number through
\begin{equation}
h_0(\mathrm{Ca}) = K_{\mathrm{Fr}} \, \ell_c \, \mathrm{Ca}^{2/3} \times \alpha_{\mathrm{Fr}} \left(\frac{\mathrm{Ma}}{\mathrm{Ca}^{2/3}} \right).
\label{eq:h_theorique}
\end{equation}
%
%
\subsection{LLD configuration}
Similarly, we propose a non-linear fit of the thickening factor $\alpha_{\mathrm{LLD}}$ defined as $h_0$ divided by the half film thickness predicted by LLD model in the stress-free limit, namely $K_{\mathrm{LLD}}^{\mathrm{sf}} \,  \ell_c \, \mathrm{Ca}^{2/3}$:
\begin{equation}
\alpha_{\mathrm{LLD}}(\Lambda) \approx \frac{1}{2} \left[ 3.603 - 1.436 \tanh \left(15.75 - 15.52 \, \Lambda^{0.030} \right) \right].
\label{eq:_alpha_Lambda_LLD}
\end{equation}
Since $K_{\mathrm{LLD}}^{\mathrm{sf}} = K_{\mathrm{Fr}}$, the theoretical film thickness as a function of the capillary number is also obtained from eq. \eqref{eq:h_theorique}, replacing the function $\alpha_{\mathrm{Frankel}}$ by $\alpha_{\mathrm{LLD}}$ for coated films. \\
%
%
\section*{Acknowledgments}
The authors would like to thank Dominique Langevin and Wiebke Drenckhan for valuable advice and comments. L.C. was supported by ANR F2F and thanks Martina Pepicelli for her help with the Langmuir trough measurments. B.S. thanks the F.R.S.-FNRS for funding. B.S. and J.V. thank the IAP-7/38 MicroMAST project for supporting this research. This work was performed under the umbrella of COST Action MP1106.
%
%
%
\footnotesize{
\bibliography{biblio} 

\providecommand*{\mcitethebibliography}{\thebibliography}
\csname @ifundefined\endcsname{endmcitethebibliography}
{\let\endmcitethebibliography\endthebibliography}{}
\begin{mcitethebibliography}{49}
\providecommand*{\natexlab}[1]{#1}
\providecommand*{\mciteSetBstSublistMode}[1]{}
\providecommand*{\mciteSetBstMaxWidthForm}[2]{}
\providecommand*{\mciteBstWouldAddEndPuncttrue}
  {\def\EndOfBibitem{\unskip.}}
\providecommand*{\mciteBstWouldAddEndPunctfalse}
  {\let\EndOfBibitem\relax}
\providecommand*{\mciteSetBstMidEndSepPunct}[3]{}
\providecommand*{\mciteSetBstSublistLabelBeginEnd}[3]{}
\providecommand*{\EndOfBibitem}{}
\mciteSetBstSublistMode{f}
\mciteSetBstMaxWidthForm{subitem}
{(\emph{\alph{mcitesubitemcount}})}
\mciteSetBstSublistLabelBeginEnd{\mcitemaxwidthsubitemform\space}
{\relax}{\relax}

\bibitem[Stevenson(2012)]{Stevenson2012}
P.~Stevenson, \emph{Foam engineering: fundamentals and applications},
  Wiley.com, 2012\relax
\mciteBstWouldAddEndPuncttrue
\mciteSetBstMidEndSepPunct{\mcitedefaultmidpunct}
{\mcitedefaultendpunct}{\mcitedefaultseppunct}\relax
\EndOfBibitem
\bibitem[Bhushan and Gupta(1991)]{Bhushan1991}
B.~Bhushan and B.~K. Gupta, \emph{Handbook of tribology: materials, coatings,
  and surface treatments}, McGraw-Hill, New York, NY (United States),
  1991\relax
\mciteBstWouldAddEndPuncttrue
\mciteSetBstMidEndSepPunct{\mcitedefaultmidpunct}
{\mcitedefaultendpunct}{\mcitedefaultseppunct}\relax
\EndOfBibitem
\bibitem[Kistler and Schweizer(1997)]{Kistler1997}
S.~F. Kistler and P.~M. Schweizer, \emph{Liquid film coating}, Springer,
  1997\relax
\mciteBstWouldAddEndPuncttrue
\mciteSetBstMidEndSepPunct{\mcitedefaultmidpunct}
{\mcitedefaultendpunct}{\mcitedefaultseppunct}\relax
\EndOfBibitem
\bibitem[Van~Nierop \emph{et~al.}(2009)Van~Nierop, Scheid, and
  Stone]{vanNieropCORR2009}
E.~A. Van~Nierop, B.~Scheid and H.~A. Stone, \emph{Journal of fluid mechanics},
  2009, \textbf{630}, 443--443\relax
\mciteBstWouldAddEndPuncttrue
\mciteSetBstMidEndSepPunct{\mcitedefaultmidpunct}
{\mcitedefaultendpunct}{\mcitedefaultseppunct}\relax
\EndOfBibitem
\bibitem[Seiwert \emph{et~al.}(2014)Seiwert, Dollet, and Cantat]{Seiwert2014}
J.~Seiwert, B.~Dollet and I.~Cantat, \emph{Journal of Fluid Mechanics}, 2014,
  \textbf{739}, 124--142\relax
\mciteBstWouldAddEndPuncttrue
\mciteSetBstMidEndSepPunct{\mcitedefaultmidpunct}
{\mcitedefaultendpunct}{\mcitedefaultseppunct}\relax
\EndOfBibitem
\bibitem[Mysels \emph{et~al.}(1959)Mysels, Frankel, and Shinoda]{Mysels1959}
K.~J. Mysels, S.~Frankel and K.~Shinoda, \emph{Soap films: studies of their
  thinning and a bibliography}, Pergamon Press, 1959\relax
\mciteBstWouldAddEndPuncttrue
\mciteSetBstMidEndSepPunct{\mcitedefaultmidpunct}
{\mcitedefaultendpunct}{\mcitedefaultseppunct}\relax
\EndOfBibitem
\bibitem[Bruinsma \emph{et~al.}(1992)Bruinsma, Di~Meglio, Qu{\'e}r{\'e}, and
  Cohen-Addad]{Bruinsma1992}
R.~Bruinsma, J.~M. Di~Meglio, D.~Qu{\'e}r{\'e} and S.~Cohen-Addad,
  \emph{Langmuir}, 1992, \textbf{8}, 3161--3167\relax
\mciteBstWouldAddEndPuncttrue
\mciteSetBstMidEndSepPunct{\mcitedefaultmidpunct}
{\mcitedefaultendpunct}{\mcitedefaultseppunct}\relax
\EndOfBibitem
\bibitem[Scheid \emph{et~al.}(2010)Scheid, van Nierop, and
  Stone]{Scheid2010_thermocapillary}
B.~Scheid, E.~A. van Nierop and H.~A. Stone, \emph{Applied physics letters},
  2010, \textbf{97}, 171906--171906\relax
\mciteBstWouldAddEndPuncttrue
\mciteSetBstMidEndSepPunct{\mcitedefaultmidpunct}
{\mcitedefaultendpunct}{\mcitedefaultseppunct}\relax
\EndOfBibitem
\bibitem[Scheid \emph{et~al.}(2012)Scheid, van Nierop, and
  Stone]{Scheid2012_thermocapillary}
B.~Scheid, E.~A. van Nierop and H.~A. Stone, \emph{Physics of fluids}, 2012,
  \textbf{24}, 032107\relax
\mciteBstWouldAddEndPuncttrue
\mciteSetBstMidEndSepPunct{\mcitedefaultmidpunct}
{\mcitedefaultendpunct}{\mcitedefaultseppunct}\relax
\EndOfBibitem
\bibitem[Landau and Levich(1942)]{Landau1942}
L.~Landau and B.~Levich, \emph{Acta Physicochim. USSR.}, 1942, \textbf{17},
  42--54\relax
\mciteBstWouldAddEndPuncttrue
\mciteSetBstMidEndSepPunct{\mcitedefaultmidpunct}
{\mcitedefaultendpunct}{\mcitedefaultseppunct}\relax
\EndOfBibitem
\bibitem[Derjaguin(1943)]{Derjaguin1943}
B.~Derjaguin, \emph{Acta Physicochim. USSR.}, 1943, \textbf{20}, 349\relax
\mciteBstWouldAddEndPuncttrue
\mciteSetBstMidEndSepPunct{\mcitedefaultmidpunct}
{\mcitedefaultendpunct}{\mcitedefaultseppunct}\relax
\EndOfBibitem
\bibitem[Sagis(2011)]{Sagis2011}
L.~M. Sagis, \emph{Reviews of Modern Physics}, 2011, \textbf{83}, 1367\relax
\mciteBstWouldAddEndPuncttrue
\mciteSetBstMidEndSepPunct{\mcitedefaultmidpunct}
{\mcitedefaultendpunct}{\mcitedefaultseppunct}\relax
\EndOfBibitem
\bibitem[Lionti-Addad and di~Meglio(1992)]{LiontiAddad1992}
S.~Lionti-Addad and J.~M. di~Meglio, \emph{Langmuir}, 1992, \textbf{8},
  324--327\relax
\mciteBstWouldAddEndPuncttrue
\mciteSetBstMidEndSepPunct{\mcitedefaultmidpunct}
{\mcitedefaultendpunct}{\mcitedefaultseppunct}\relax
\EndOfBibitem
\bibitem[Lal and di~Meglio(1994)]{Lal1994}
J.~Lal and J.-M. di~Meglio, \emph{Journal of colloid and interface science},
  1994, \textbf{164}, 506--509\relax
\mciteBstWouldAddEndPuncttrue
\mciteSetBstMidEndSepPunct{\mcitedefaultmidpunct}
{\mcitedefaultendpunct}{\mcitedefaultseppunct}\relax
\EndOfBibitem
\bibitem[Adelizzi and Troian(2004)]{Adelizzi2004}
E.~A. Adelizzi and S.~M. Troian, \emph{Langmuir}, 2004, \textbf{20},
  7482--7492\relax
\mciteBstWouldAddEndPuncttrue
\mciteSetBstMidEndSepPunct{\mcitedefaultmidpunct}
{\mcitedefaultendpunct}{\mcitedefaultseppunct}\relax
\EndOfBibitem
\bibitem[Berg \emph{et~al.}(2005)Berg, Adelizzi, and Troian]{Berg2005}
S.~Berg, E.~A. Adelizzi and S.~M. Troian, \emph{Langmuir}, 2005, \textbf{21},
  3867--3876\relax
\mciteBstWouldAddEndPuncttrue
\mciteSetBstMidEndSepPunct{\mcitedefaultmidpunct}
{\mcitedefaultendpunct}{\mcitedefaultseppunct}\relax
\EndOfBibitem
\bibitem[Saulnier \emph{et~al.}(2011)Saulnier, Restagno, Delacotte, Langevin,
  and Rio]{Saulnier2011}
L.~Saulnier, F.~Restagno, J.~Delacotte, D.~Langevin and E.~Rio,
  \emph{Langmuir}, 2011, \textbf{27}, 13406--13409\relax
\mciteBstWouldAddEndPuncttrue
\mciteSetBstMidEndSepPunct{\mcitedefaultmidpunct}
{\mcitedefaultendpunct}{\mcitedefaultseppunct}\relax
\EndOfBibitem
\bibitem[Morey(1940)]{Morey1940}
F.~C. Morey, \emph{J. Res. Nat. Bur. Stand.}, 1940, \textbf{25}, 385--393\relax
\mciteBstWouldAddEndPuncttrue
\mciteSetBstMidEndSepPunct{\mcitedefaultmidpunct}
{\mcitedefaultendpunct}{\mcitedefaultseppunct}\relax
\EndOfBibitem
\bibitem[Homsy and Krechetnikov(2005)]{Krechetnikov2005}
R.~Homsy and G.~M. Krechetnikov, \emph{Physics of Fluids}, 2005, \textbf{17},
  102108\relax
\mciteBstWouldAddEndPuncttrue
\mciteSetBstMidEndSepPunct{\mcitedefaultmidpunct}
{\mcitedefaultendpunct}{\mcitedefaultseppunct}\relax
\EndOfBibitem
\bibitem[Snoeijer \emph{et~al.}(2008)Snoeijer, Ziegler, Andreotti, Fermigier,
  and Eggers]{Snoeijer2008}
J.~Snoeijer, J.~Ziegler, B.~Andreotti, M.~Fermigier and J.~Eggers,
  \emph{Physical Review Letters}, 2008, \textbf{100}, 244502\relax
\mciteBstWouldAddEndPuncttrue
\mciteSetBstMidEndSepPunct{\mcitedefaultmidpunct}
{\mcitedefaultendpunct}{\mcitedefaultseppunct}\relax
\EndOfBibitem
\bibitem[Qu{\'e}r{\'e}(1999)]{Quere1999}
D.~Qu{\'e}r{\'e}, \emph{Annual Review of Fluid Mechanics}, 1999, \textbf{31},
  347--384\relax
\mciteBstWouldAddEndPuncttrue
\mciteSetBstMidEndSepPunct{\mcitedefaultmidpunct}
{\mcitedefaultendpunct}{\mcitedefaultseppunct}\relax
\EndOfBibitem
\bibitem[Shen \emph{et~al.}(2002)Shen, Gleason, McKinley, and Stone]{Shen2002}
A.~Q. Shen, B.~Gleason, G.~H. McKinley and H.~A. Stone, \emph{Physics of
  Fluids}, 2002, \textbf{14}, 4055--4068\relax
\mciteBstWouldAddEndPuncttrue
\mciteSetBstMidEndSepPunct{\mcitedefaultmidpunct}
{\mcitedefaultendpunct}{\mcitedefaultseppunct}\relax
\EndOfBibitem
\bibitem[Bretherton(1961)]{Bretherton1961}
F.~Bretherton, \emph{J. Fluid Mech}, 1961, \textbf{10}, 166--188\relax
\mciteBstWouldAddEndPuncttrue
\mciteSetBstMidEndSepPunct{\mcitedefaultmidpunct}
{\mcitedefaultendpunct}{\mcitedefaultseppunct}\relax
\EndOfBibitem
\bibitem[Van~Nierop \emph{et~al.}(2008)Van~Nierop, Scheid, and
  Stone]{vanNierop2008}
E.~A. Van~Nierop, B.~Scheid and H.~A. Stone, \emph{Journal of fluid mechanics},
  2008, \textbf{602}, 119\relax
\mciteBstWouldAddEndPuncttrue
\mciteSetBstMidEndSepPunct{\mcitedefaultmidpunct}
{\mcitedefaultendpunct}{\mcitedefaultseppunct}\relax
\EndOfBibitem
\bibitem[van Nierop \emph{et~al.}(2009)van Nierop, Keupp, and
  Stone]{vanNierop2009_formation}
E.~van Nierop, D.~Keupp and H.~Stone, \emph{Europhysics Letters}, 2009,
  \textbf{88}, 66005\relax
\mciteBstWouldAddEndPuncttrue
\mciteSetBstMidEndSepPunct{\mcitedefaultmidpunct}
{\mcitedefaultendpunct}{\mcitedefaultseppunct}\relax
\EndOfBibitem
\bibitem[Delacotte \emph{et~al.}(2012)Delacotte, Montel, Restagno, Scheid,
  Dollet, Stone, Langevin, and Rio]{Delacotte2012}
J.~Delacotte, L.~Montel, F.~Restagno, B.~Scheid, B.~Dollet, H.~A. Stone,
  D.~Langevin and E.~Rio, \emph{Langmuir}, 2012, \textbf{28}, 3821--3830\relax
\mciteBstWouldAddEndPuncttrue
\mciteSetBstMidEndSepPunct{\mcitedefaultmidpunct}
{\mcitedefaultendpunct}{\mcitedefaultseppunct}\relax
\EndOfBibitem
\bibitem[Park(1991)]{Park1991}
C.-W. Park, \emph{Journal of colloid and interface science}, 1991,
  \textbf{146}, 382--394\relax
\mciteBstWouldAddEndPuncttrue
\mciteSetBstMidEndSepPunct{\mcitedefaultmidpunct}
{\mcitedefaultendpunct}{\mcitedefaultseppunct}\relax
\EndOfBibitem
\bibitem[Tiwari and Davis(2006)]{Tiwari2006}
N.~Tiwari and J.~M. Davis, \emph{Physics of Fluids}, 2006, \textbf{18},
  022102\relax
\mciteBstWouldAddEndPuncttrue
\mciteSetBstMidEndSepPunct{\mcitedefaultmidpunct}
{\mcitedefaultendpunct}{\mcitedefaultseppunct}\relax
\EndOfBibitem
\bibitem[Campana \emph{et~al.}(2010)Campana, Ubal, Giavedoni, and
  Saita]{Campana2010}
D.~M. Campana, S.~Ubal, M.~D. Giavedoni and F.~A. Saita, \emph{Physics of
  Fluids}, 2010, \textbf{22}, 032103\relax
\mciteBstWouldAddEndPuncttrue
\mciteSetBstMidEndSepPunct{\mcitedefaultmidpunct}
{\mcitedefaultendpunct}{\mcitedefaultseppunct}\relax
\EndOfBibitem
\bibitem[Campana \emph{et~al.}(2011)Campana, Ubal, Giavedoni, and
  Saita]{Campana2011}
D.~M. Campana, S.~Ubal, M.~D. Giavedoni and F.~A. Saita, \emph{Physics of
  Fluids (1994-present)}, 2011, \textbf{23}, 052102\relax
\mciteBstWouldAddEndPuncttrue
\mciteSetBstMidEndSepPunct{\mcitedefaultmidpunct}
{\mcitedefaultendpunct}{\mcitedefaultseppunct}\relax
\EndOfBibitem
\bibitem[Schwartz and Roy(1999)]{Schwartz1999}
L.~Schwartz and R.~Roy, \emph{Journal of colloid and interface science}, 1999,
  \textbf{218}, 309--323\relax
\mciteBstWouldAddEndPuncttrue
\mciteSetBstMidEndSepPunct{\mcitedefaultmidpunct}
{\mcitedefaultendpunct}{\mcitedefaultseppunct}\relax
\EndOfBibitem
\bibitem[Naire \emph{et~al.}(2001)Naire, Braun, and Snow]{Naire2001}
S.~Naire, R.~Braun and S.~Snow, \emph{Physics of Fluids}, 2001, \textbf{13},
  2492\relax
\mciteBstWouldAddEndPuncttrue
\mciteSetBstMidEndSepPunct{\mcitedefaultmidpunct}
{\mcitedefaultendpunct}{\mcitedefaultseppunct}\relax
\EndOfBibitem
\bibitem[Scheid \emph{et~al.}(2010)Scheid, Delacotte, Dollet, Rio, Restagno,
  Van~Nierop, Cantat, Langevin, and Stone]{Scheid2010_LLD}
B.~Scheid, J.~Delacotte, B.~Dollet, E.~Rio, F.~Restagno, E.~Van~Nierop,
  I.~Cantat, D.~Langevin and H.~A. Stone, \emph{Europhysics Letters}, 2010,
  \textbf{90}, 24002\relax
\mciteBstWouldAddEndPuncttrue
\mciteSetBstMidEndSepPunct{\mcitedefaultmidpunct}
{\mcitedefaultendpunct}{\mcitedefaultseppunct}\relax
\EndOfBibitem
\bibitem[Bhamla \emph{et~al.}(2014)Bhamla, Giacomin, Balemans, and
  Fuller]{Bhamla2014}
M.~S. Bhamla, C.~E. Giacomin, C.~Balemans and G.~G. Fuller, \emph{Soft Matter},
  2014\relax
\mciteBstWouldAddEndPuncttrue
\mciteSetBstMidEndSepPunct{\mcitedefaultmidpunct}
{\mcitedefaultendpunct}{\mcitedefaultseppunct}\relax
\EndOfBibitem
\bibitem[Sonin \emph{et~al.}(1994)Sonin, Bonfillon, and Langevin]{Sonin1994}
A.~Sonin, A.~Bonfillon and D.~Langevin, \emph{Journal of colloid and interface
  science}, 1994, \textbf{162}, 323--330\relax
\mciteBstWouldAddEndPuncttrue
\mciteSetBstMidEndSepPunct{\mcitedefaultmidpunct}
{\mcitedefaultendpunct}{\mcitedefaultseppunct}\relax
\EndOfBibitem
\bibitem[Prins \emph{et~al.}(1967)Prins, Arcuri, and Van Den~Tempel]{Prins1967}
A.~Prins, C.~Arcuri and M.~Van Den~Tempel, \emph{Journal of Colloid and
  Interface Science}, 1967, \textbf{24}, 84--90\relax
\mciteBstWouldAddEndPuncttrue
\mciteSetBstMidEndSepPunct{\mcitedefaultmidpunct}
{\mcitedefaultendpunct}{\mcitedefaultseppunct}\relax
\EndOfBibitem
\bibitem[Lucassen(1981)]{Lucassen1981}
J.~Lucassen, \emph{Anionic surfactants: physical chemistry of surfactant
  action}, 1981, \textbf{11}, 217--265\relax
\mciteBstWouldAddEndPuncttrue
\mciteSetBstMidEndSepPunct{\mcitedefaultmidpunct}
{\mcitedefaultendpunct}{\mcitedefaultseppunct}\relax
\EndOfBibitem
\bibitem[Gibbs(1906)]{Gibbs1906}
J.~W. Gibbs, \emph{The scientific papers of J. Willard Gibbs}, Longmans, Green
  and Company, 1906, vol.~1\relax
\mciteBstWouldAddEndPuncttrue
\mciteSetBstMidEndSepPunct{\mcitedefaultmidpunct}
{\mcitedefaultendpunct}{\mcitedefaultseppunct}\relax
\EndOfBibitem
\bibitem[Saulnier(2012)]{Saulnier_phd}
L.~Saulnier, \emph{PhD thesis}, Universit\'{e} Paris-Sud, 2012\relax
\mciteBstWouldAddEndPuncttrue
\mciteSetBstMidEndSepPunct{\mcitedefaultmidpunct}
{\mcitedefaultendpunct}{\mcitedefaultseppunct}\relax
\EndOfBibitem
\bibitem[Howell(1996)]{Howell1996}
P.~Howell, \emph{PhD thesis}, Oxford University, 1996\relax
\mciteBstWouldAddEndPuncttrue
\mciteSetBstMidEndSepPunct{\mcitedefaultmidpunct}
{\mcitedefaultendpunct}{\mcitedefaultseppunct}\relax
\EndOfBibitem
\bibitem[Stebe and Barthes-Biesel(1995)]{Stebe1995}
K.~Stebe and D.~Barthes-Biesel, \emph{Journal of Fluid Mechanics}, 1995,
  \textbf{286}, 25--48\relax
\mciteBstWouldAddEndPuncttrue
\mciteSetBstMidEndSepPunct{\mcitedefaultmidpunct}
{\mcitedefaultendpunct}{\mcitedefaultseppunct}\relax
\EndOfBibitem
\bibitem[Bendure(1971)]{Bendure1971}
R.~L. Bendure, \emph{Journal of Colloid and Interface Science}, 1971,
  \textbf{35}, 238--248\relax
\mciteBstWouldAddEndPuncttrue
\mciteSetBstMidEndSepPunct{\mcitedefaultmidpunct}
{\mcitedefaultendpunct}{\mcitedefaultseppunct}\relax
\EndOfBibitem
\bibitem[Lin \emph{et~al.}(2002)Lin, Lee, and Shao]{Lin2002}
S.-Y. Lin, Y.-C. Lee and M.-J. Shao, \emph{Journal of the Chinese Institute of
  Chemical Engineers}, 2002, \textbf{33}, 631--643\relax
\mciteBstWouldAddEndPuncttrue
\mciteSetBstMidEndSepPunct{\mcitedefaultmidpunct}
{\mcitedefaultendpunct}{\mcitedefaultseppunct}\relax
\EndOfBibitem
\bibitem[Doedel \emph{et~al.}(2007)Doedel, Paffenroth, Champneys, Fairgrieve,
  Kuznetsov, Oldeman, Sandstede, and Wang]{AUTO07P}
E.~Doedel, R.~Paffenroth, A.~Champneys, T.~Fairgrieve, Y.~A. Kuznetsov,
  B.~Oldeman, B.~Sandstede and X.~Wang, \emph{AUTO-07P: Continuation and
  bifurcation software for ordinary differential equations}, 2007, Available
  for download from http://indy. cs. concordia. ca/auto\relax
\mciteBstWouldAddEndPuncttrue
\mciteSetBstMidEndSepPunct{\mcitedefaultmidpunct}
{\mcitedefaultendpunct}{\mcitedefaultseppunct}\relax
\EndOfBibitem
\bibitem[Saulnier \emph{et~al.}(2014)Saulnier, Champougny, Bastien, Restagno,
  Langevin, and Rio]{Saulnier2014}
L.~Saulnier, L.~Champougny, G.~Bastien, F.~Restagno, D.~Langevin and E.~Rio,
  \emph{Soft Matter}, 2014\relax
\mciteBstWouldAddEndPuncttrue
\mciteSetBstMidEndSepPunct{\mcitedefaultmidpunct}
{\mcitedefaultendpunct}{\mcitedefaultseppunct}\relax
\EndOfBibitem
\bibitem[Adami and Caps(2013)]{Caps2013}
N.~Adami and H.~Caps, arXiv:1310.0454 [physics.flu-dyn]\relax
\mciteBstWouldAddEndPuncttrue
\mciteSetBstMidEndSepPunct{\mcitedefaultmidpunct}
{\mcitedefaultendpunct}{\mcitedefaultseppunct}\relax
\EndOfBibitem
\bibitem[Lucassen-Reynders \emph{et~al.}(2001)Lucassen-Reynders, Cagna, and
  Lucassen]{Lucassen2001}
E.~Lucassen-Reynders, A.~Cagna and J.~Lucassen, \emph{Colloids and Surfaces A},
  2001, \textbf{186}, 63--72\relax
\mciteBstWouldAddEndPuncttrue
\mciteSetBstMidEndSepPunct{\mcitedefaultmidpunct}
{\mcitedefaultendpunct}{\mcitedefaultseppunct}\relax
\EndOfBibitem
\bibitem[Chang and Franses(1995)]{Chang1995}
C.-H. Chang and E.~I. Franses, \emph{Colloids and Surfaces A: Physicochemical
  and Engineering Aspects}, 1995, \textbf{100}, 1--45\relax
\mciteBstWouldAddEndPuncttrue
\mciteSetBstMidEndSepPunct{\mcitedefaultmidpunct}
{\mcitedefaultendpunct}{\mcitedefaultseppunct}\relax
\EndOfBibitem
\bibitem[Ou~Ramdane and Qu{\'e}r{\'e}(1997)]{OuRamdane1997}
O.~Ou~Ramdane and D.~Qu{\'e}r{\'e}, \emph{Langmuir}, 1997, \textbf{13},
  2911--2916\relax
\mciteBstWouldAddEndPuncttrue
\mciteSetBstMidEndSepPunct{\mcitedefaultmidpunct}
{\mcitedefaultendpunct}{\mcitedefaultseppunct}\relax
\EndOfBibitem
\end{mcitethebibliography}
\bibliographystyle{rsc} 
}
%
\end{document}